\documentclass[letterpaper]{article} 
\usepackage{aaai24}  
\usepackage{times}  
\usepackage{helvet}  
\usepackage{courier}  
\usepackage[hyphens]{url}  
\usepackage{graphicx} 
\usepackage{natbib}  
\usepackage{caption} 
\usepackage{algorithm}
\usepackage{algorithmicx}
\usepackage{algpseudocode}

\usepackage{newfloat}
\usepackage{listings}
\DeclareCaptionStyle{ruled}{labelfont=normalfont,labelsep=colon,strut=off} 
\floatstyle{ruled}
\newfloat{listing}{tb}{lst}{}
\floatname{listing}{Listing}
\title{Nowcasting Temporal Trends Using Indirect Surveys}
\author{
    Ajitesh Srivastava\textsuperscript{\rm 1}, 
    Juan Marcos Ramírez\textsuperscript{\rm 2}, Sergio Díaz-Aranda\textsuperscript{\rm 2,3}, Jose Aguilar\textsuperscript{\rm 2}, Antonio Ortega\textsuperscript{\rm 1}, Antonio {Fernández Anta}\textsuperscript{\rm 2}, Rosa Elvira Lillo\textsuperscript{\rm 3}
}
\affiliations{
    \textsuperscript{\rm 1} University of Southern California, USA\\
    \textsuperscript{\rm 2} IMDEA Networks Institute, Spain\\
    \textsuperscript{\rm 3} Universidad Carlos III, Spain\\
    
    ajiteshs@usc.edu, juan.ramirez@imdea.org, sergio.diaz@imdea.org, jose.aguilar@imdea.org, aortega@usc.edu, antonio.fernandez@imdea.org, lillo@est-econ.uc3m.es
}

\usepackage{multirow}
\usepackage{enumitem}
\usepackage{amssymb}
\usepackage{amsfonts}
\usepackage{amsthm}
\usepackage{amsmath}

\newtheorem{assumption}{Assumption}
\newtheorem{lemma}{Lemma}
\newtheorem{theorem}{Theorem}
\newtheorem{observation}{Observation}
\newtheorem{definition}{Definition}

\newcommand{\remove}[1]{{}}

\begin{document}

\maketitle

\begin{abstract}
Indirect surveys, in which respondents provide information about other people they know, have been proposed for estimating (nowcasting) the size of a \emph{hidden population} where privacy is important or the hidden population is hard to reach. 
Examples include estimating casualties in an earthquake, conditions among female sex workers, and the prevalence of drug use and infectious diseases.
The Network Scale-up Method (NSUM) is the classical approach to developing estimates from indirect surveys, but it was designed for one-shot surveys. Further, it requires certain assumptions and asking for or estimating the number of individuals in each respondent's network.
In recent years, surveys have been increasingly deployed online and can collect data continuously (e.g., COVID-19 surveys on Facebook during much of the pandemic). Conventional NSUM can be applied to these scenarios by analyzing the data independently at each point in time, but this misses the opportunity of leveraging the temporal dimension. 
We propose to use the responses from indirect surveys collected over time and develop analytical tools (i)  to prove that indirect surveys can provide better estimates for the trends of the hidden population over time, as compared to direct surveys and (ii) to identify appropriate temporal aggregations to improve the estimates. We demonstrate through extensive simulations that our approach outperforms traditional NSUM and direct surveying methods. 
We also empirically demonstrate the superiority of our approach on a real indirect survey dataset of COVID-19 cases.
\end{abstract}




\section{Introduction}

Direct reporting through surveys is the most widely used method for collecting data on a given characteristic among individuals in a population. It is well known that direct surveys sometimes face reliability and efficiency problems, as respondents may refuse to participate or may choose to misreport sensitive private information. 
An alternative approach to overcome some of these issues is using surveys with \textit{indirect} reporting, where respondents answer questions about people they know instead of providing information about themselves. These surveys collect what is known in the literature as \textit{aggregated relational data (ARD)}. 
Their main advantages are primarily two: (1) privacy is preserved as the respondents do not have to report their own status, thus improving participation and data collection about sensitive populations \cite{rossier2010anonymous}; (2)  one individual response gives the researchers access to information about many different individuals, thus leading to cost reductions in the data collection \cite{breza2020using}, \cite{alix2021better}.  
Indirect surveys have been employed in a variety of domains, such as estimating the number of casualties in an earthquake \cite{bernard1989estimating},  conditions among female sex workers \cite{jing2018combining}, or the prevalence of drug use \cite{salganik2010estimating}, HIV \cite{teo2019estimating} or COVID-19 \cite{garcia2021estimating}. 

While indirect surveys have a long history~\cite{laga2021thirty}, the ubiquity of internet access among the general public has made it possible to develop \textit{online indirect surveys}, which can be deployed rapidly and allow the continuous collection of ARD, instead of consisting of one-shot surveys.
%
The importance of recurrent online surveys, including indirect surveys, has become apparent during the COVID-19 pandemic, where 
one of the most important challenges has been 
estimating (nowcasting) the number of cases (especially when testing was not widely available), the number of deaths, the number of people vaccinated, etc. 
While the best-known online surveys are the COVID-19 Trends and Impact Surveys (CTIS)~\cite{astley2021global,salomon2021us}, 
which collected more than 100,000 responses daily, other surveys were also deployed  \cite{geldsetzer2020knowledge,oliver2020assessing,garcia2021estimating}. 

Our main goals are  (i) to \textit{quantify the advantages of indirect surveys over direct} ones for nowcasting, by determining under what conditions, more accurate estimates can be obtained via indirect questions; and (ii) to  \textit{develop a new method to identify a hidden temporal trend} from the continuously collected ARD from indirect surveys.
To achieve these goals, we combine a detailed theoretical analysis with extensive experimental evaluations with both synthetic and real datasets. Given the availability, throughout the paper, we use the COVID-19 surveys dataset as a test case. 

\subsection{Related Work}

ARD from indirect surveys are used to nowcast the size of a subpopulation or \textit{hidden population}, 
i.e., the fraction of the population that has some characteristic. For example, in a COVID-19-related survey, a respondent may report on how many people they know who have tested positive recently, and this information will be used to estimate the fraction of the overall population that is infected at that time.  
Thus, each respondent is expected to provide information about their own \textit{personal network} (PN) -- how many people they know-- \textit{and} the number of people they know who are part of the hidden population 
(e.g., tested positive). 

\subsubsection{NSUM} The methods proposed in the literature for the estimation of the size of hidden populations (people infected in our example) using ARD are generally known as Network Scale-Up Methods (NSUM) \cite{laga2021thirty}.
To estimate the  size of the hidden population from the ARD,  
NSUM assumes that the proportion of those belonging to the hidden population within a respondent's PN is a good approximation to the same proportion in the overall population. 
NSUM can work well under several assumptions: (1) the respondents can accurately recall the people in their PN, (2) the respondents know, for each person in their PN, if they belong to the hidden population, and (3) all individuals have the same probability of belonging to the hidden population. Errors resulting from  violations of these conditions are called recall error, transmission error, and barrier effects, respectively (see \cite{laga2021thirty} for more details). 
Multiple NSUM extensions have been  proposed~\cite{laga2021thirty}
but all of them require to request or estimate PN sizes \cite{killworth1998estimation,laga2021thirty,garcia2021estimating}, or ask individuals in the hidden population about those who know their condition \cite{feehan2016generalizing}. 

Our goal is to obtain better estimates of the evolution of the hidden population size \textit{without a need to obtain or estimate the size of individuals' PNs} and, thus, without having to rely on the above recall assumptions. Specifically, one of the key contributions of our work is to leverage the temporal dynamics to improve estimates from continuously collected data. We provide analytical tools to unveil the advantages of appropriately aggregating continuous indirect surveys rather than simply using existing methods over fixed temporal windows. 
To the best of our knowledge, prior work does not 
leverage the temporal nature of continuous surveys.

\subsection{Contributions}

We propose a method to estimate the evolution of the size of the hidden population over a period of time using indirect surveys. 
Our contributions are as follows:
\begin{itemize}[leftmargin=*]
    \item We propose a latent graph formulation that allows us to prove that the expected response to the indirect survey is proportional to the size of the hidden population (Theorem~\ref{thm:prop}). Unlike existing work, we do not assume that every individual has the same probability of reporting someone belonging to the hidden population.
    \item We prove that within a reasonable upper bound on the latent graph degree variance, the indirect survey provides a better estimate of the hidden population than the direct survey given the same number of samples (Theorem~\ref{thm:indirect_better}).
    \item We leverage the smoothness of the underlying temporal dynamics to show that a weighted moving average provides better estimates than a series of individual estimates 
    (Lemma~\ref{lem:extra-dev},  Theorem~\ref{thm:accum}, and Theorem~\ref{thm:accum2}).
    \item We verify our claims through a simulated generation of the hidden population with a dynamic process and a simulated survey. We present the impact of various survey parameters 
    (Section~\ref{sec:simul}).
    \item We evaluate our approach in the estimation of COVID-19 cases in the US for a period of 18 months (Section~\ref{sec:CTIS}).
\end{itemize}
Our analytical results can be useful for survey design.
Note that our objective is nowcasting (rather than forecasting) the time series of the size of the hidden population.

\section{Methodology}
\subsection{Latent Graph Formulation}

Consider a population given by a set $N$. Suppose that, at time $t$, $N$ contains a hidden population denoted by $H_t  \subseteq N$, leading to a hidden population rate $f_t = |H_t|/|N|$. Let $G=(N, E)$ be a directed graph, where $N$ is the set of nodes and $E$ is the set of edges. In particular, $G$ includes an edge $(v,u)$ if node $u$ possesses knowledge of node $v$ and is willing to report whether node $v$ belongs to the hidden population. 
Also, we allow self-loops $(u, u)$. These edges may not be the same as those in the contact graph or the social network graph containing the same nodes. The edges present on this (unobserved) graph depend on the specific wording of the survey questions, e.g., ``how many in your community ...", ``... your household'',  ``your immediate neighbors and coworkers'', etc.


Consider a random process that selects \textbf{one node}, at time $t$ to report the number of its neighbors belonging to the hidden population.  We denote by $X_t$ the random variable corresponding to this response. 
In the surveys, at a given time $t$, multiple nodes are selected randomly (possibly, with replacement) that provide multiple observations for $X_t$.
We will later use the mean and variance of $X_t$ to identify the properties of the \textit{sample mean} $\bar{X}_t$ obtained from these responses. 
Finally, $D$ is a random variable representing the in-degree of a randomly selected node, with $\mathbb{E}(D) = \mu_D$.

\begin{assumption}\label{assume:indep}
For any node $v$ belonging to the neighborhood of node $u$, the event $v \in H_t$ is independent of the in-degree of $u$. 
\end{assumption}

This implies that \textit{\textbf{having a certain in-degree does not affect whether a randomly selected neighbor is part of the hidden population}}. This assumption is more flexible than that used in the traditional NSUM approach, in which every node must have the same probability of finding a neighbor belonging to the hidden population \cite{laga2021thirty}. 
Under Assumption~\ref{assume:indep},  nodes can have different probabilities of having neighbors in $H_t$ (for example, this could depend on their respective occupations). 
However, if we consider the union of neighbors of all nodes with a particular indegree, the probability of a random node in this union belonging to $H_t$ remains $f_t$, irrespective of the indegree considered.
This assumption will allow us to eliminate the need to ask for the in-degree of each node. 
Based on Assumption~\ref{assume:indep}, we have the following theorem.


\begin{theorem}\label{thm:prop}
$\mathbb{E}(X_t) = \mu_D \cdot f_t$.
\end{theorem}

Therefore, the \textit{\textbf{mean indirect response is proportional to what we wish to estimate}}. Additionally, we make the following observation regarding the underlying graph $G$.

\begin{observation}\label{assume:mu_const}
The mean in-degree of all nodes, $\mu_D$,  remains constant over time.
\end{observation}

This observation is supported by the studies of Dunbar~\cite{dunbar2010many}. It can also be observed in the data collected by the Carnegie Mellon University US COVID-19 Trends and Impact Survey (CMU-CTIS)~\cite{salomon2021us}, where, over time, different respondents provided the household size in which they reported the number of infections 
(see Supplementary Material in full version \cite{srivastava2023estimating}).

Recall that $G$ is not an acquaintance or physical contact network. Instead, the connection of a node in $G$ represents the network of people the respondent will think of when answering the question. Respondents may not report on the same people each time the survey is completed. So \textbf{$G$ may be different at each time $t$, but the mean of the in-degrees remains constant}. 
From Observation~\ref{assume:mu_const} and Theorem~\ref{thm:prop}, the time series $\mathbb{E}(X_t)$ is proportional to time series $f_t$, representing the fraction of the hidden population, with $\mu_D$ as the constant of proportionality. 
Hence, we can estimate the trend of $f_t$ without knowing $\mu_D$. 
For applications in which precise $f_t$ values are needed, if the true value of $f_t$ is available for some $t=\tau$ then we can estimate $\mu_D = \mathbb{E}(X_{\tau})/f_{\tau}$ and use this constant to estimate $f_t$ at any $t$. For example, when $f_t$ represents the rate of active infections, 
$\tau$ could correspond to those dates for which serological studies or wastewater concentration data are available. 

\subsubsection{Comparison Against Direct Reporting.}
With direct reporting, each node reports whether it belongs to the hidden population. Thus, for a randomly selected node $v$, the response is the binary indicator function $I_v$, where $I_{v} = 1$ iff $v \in H_t$. 
Let $Y_t$ be a random variable denoting the response of a randomly selected node. Observe that $Y_t$ is a binary random variable whose samples follow a Bernoulli distribution with mean $\mathbb{E}(Y_t) = f_t$ and variance $\sigma_{Y_t}^2 = f_t(1-f_t)$. 
To compare direct and indirect reporting scenarios, we also need to compute the variance of $X_t$.
Since the links in our latent graph do not represent physical contact, neighbors of node $u$ are not necessarily dependent, and we 
introduce a parameter $\phi_t$ that controls the level of covariance. 
\begin{definition}\label{asmpt:phi}
For a pair of nodes $v_1$ and $v_2$, with a common neighbor u, $\mathbb{E}(I_{v_1}I_{v_2}|\delta(u)) = \mathbb{E}(I_{v_1}I_{v_2}) = \phi_t f_t$\,, for some $0 \leq \phi_t \leq 1$.  
\end{definition}
Here $\phi_t = f_t$ implies independence, $\phi_t < f_t$ leads to negative covariance and $\phi_t > f_t$ leads to positive covariance. Now we can find bounds on the variance of $X_t$.

\begin{lemma}\label{lem:sigma_X}
If $\sigma_D^2$ is the variance of the degree distribution,
    \begin{equation}
    \sigma_{X_t}^2 =  f_t (\mu_D^2 (\phi_t-f_t) + \mu_D (1-\phi_t) + \sigma_D^2 \phi_t)\,.
\end{equation}
    Further, $ \mu_D f_t (1-\mu_D f_t) \leq \sigma_{X_t}^2 \leq f_t ( \sigma_D^2 + \mu_D^2 (1-f_t)) $.
\end{lemma}
 
Suppose we have the same number of responses $n$ for $X_t$ and $Y_t$, and $n \gg 1$, we show that within a practical upper bound of degree variance $\sigma_D^2$, the indirect survey is a better estimator than the direct survey.

\begin{lemma} [Central Limit Theorem, CLT]\label{assume:clt}
    When the number of samples $n$ is large,  $\left(\frac{\bar{X}_t - \mathbb{E}(X_t)}{\sigma_{X_t}/\sqrt{n}}\right)$ and $\left(\frac{\bar{Y}_t - \mathbb{E}(Y_t)}{\sigma_{Y_t}/\sqrt{n}}\right)$ follow standard normal distribution, where
    $\bar{X}_t$ and $\bar{Y}_t$ are the sample means.
\end{lemma}

Now, we can show, under Lemma~\ref{assume:clt}, that the probability of deviating from the true fraction of hidden population $f_t$ is lower for the estimate obtained from indirect responses $\bar{X}_t/\mu_D$ compared to direct responses $\bar{Y}_t$.
\begin{theorem}\label{thm:indirect_better}
    For any $\lambda > 0$, $P(|\bar{X}_t/\mu_D - f_t| > \lambda) \leq P(|\bar{Y}_t - f_t| > \lambda)$, if the variance of degree distribution $\sigma_D^2 \leq \mu_D (\mu_D - 1)(1-\phi_t)/\phi_t$ 
\end{theorem}

This means \textit{\textbf{within realistic bounds on degree variance, indirect surveys are better than direct surveys.}}
The condition on $\sigma_D^2$  is reasonable for applications where $f_t$ at any given time is a small fraction of the population.  
To see this, first note that if the membership of two neighbors in the hidden population is independent, $\phi_t = f_t$. 
Assuming that the maximum degree in the graph is $\delta_{max} = 50$, and $\mu_D = 10$, the variance can be bounded using~\cite{bhatia2000better} by $\sigma_{D}^2 \leq (\delta_{max} - \mu_D)(\mu_D - 1)$.
To satisfy this bound, and still violate the assumption on degree variance in Theorem~\ref{thm:indirect_better}, would require $f_t \geq 0.2$, i.e., for $20\%$ of the population to be in the hidden population (e.g., positive with COVID-19) simultaneously. This is unrealistically high, noting that the highest number of COVID-19 tests performed (which is much higher than reported positive cases) in a week in California was approximately 11.2\% ($<$ 20\%) of the population. 
Theorem~\ref{thm:indirect_better} also motivates framing the indirect survey questions in such a way that the variance of the graph is small. 

\subsection{Leveraging Smoothness}

The hidden population in many real-life processes is \textbf{driven by smooth dynamic processes}, leading to the following. 
\begin{observation}
    \label{assume:smooth}
    $|\Delta f_t| \leq \epsilon_{f, 1} f_t$ and 
    $|\Delta^2 f_t| \leq \epsilon_{f, 2} f_t$, for some small $\epsilon_{f, 1}, \epsilon_{f, 2} \geq 0$.
\end{observation}

\begin{figure}[!t]
    \centering
    \includegraphics[width=0.9\columnwidth]{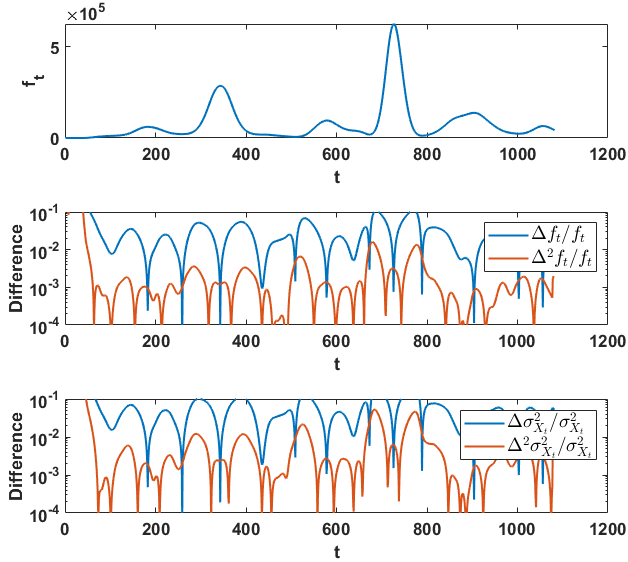}
    \caption{The first and second differences for $f_t$ and $\sigma_t$ are small -- derived from denoised COVID-19 cases in California. $t$ is the number of days since January 23, 2020.}
    \label{fig:ca_diffs}
\end{figure}

This is reasonable for epidemics as the epidemiology follows smooth dynamics which over a large population should produce smooth case counts. The reported data may appear to be noisy due to reporting behavior, schedule, and delayed dumps (a death occurring today may be reported 1-2 weeks from now). However, the artifacts of reporting to the state dashboards are not something we wish to capture. Instead, we wish to capture the actual incidence of cases based on surveys. An individual will know if their friends are infected independently of how and when it is reported to the state dashboards. 
To support this observation, we compute $f_t$ from values for COVID-19 reported cases (after denoising and removing outliers) in California based on the number of people who were reported positive within the previous 7 days of $t$.
Figure~\ref{fig:ca_diffs} shows the value of $|\Delta f_t|/f_t$ and $\Delta^2 |f_t|/f_t$ over time. Note that $|\Delta^2 f_t|/f_t \ll |\Delta f_t|/f_t \ll 1$. The same is observed for a simulated epidemic (see Section~\ref{sec:simul} for simulation details). 
We repeat the analysis for $\sigma_{X_t}^2$. 
We calculated $\sigma_{X_t}$ by setting 
 $\mu_D = 15, \sigma_D^2 = 100, \phi_t = f_t$. 
These smoothness properties are used to derive our results that demonstrate that a weighted smoothing of responses provides better estimates of $f_t$ compared to unsmoothed estimation using $\bar{X}_{t}$.

Now, we present two results (Lemmas~\ref{lem:first_diff} and~\ref{lem:second_diff}) that help \textit{ \textbf{bound the result of aggregating over smooth sequences}}.
\begin{lemma}\label{lem:first_diff}
    For any non-negative sequence $g_t$ such that $|\Delta g_t| \leq \epsilon_{g, 1} \cdot g_t, \forall t$ for some $\epsilon_{g, 1} \geq 0$, let $g_{max} = \max\{g_t, g_{t+1}, \dots, g_{t+j}\}$. Then $\left|g_{t+j} - g_t\right| \leq \frac{|j|\epsilon_{g, 1}}{1 - |j|\epsilon_{g, 1}} g_t\, \forall j\in \mathbb{Z}$.
\end{lemma}
 This implies that a time-series with bounded differences will not change rapidly in a small window of time. 
\begin{lemma}\label{lem:second_diff}
    For any non-negative sequence $g_t$ such that $|\Delta^2 g_t| \leq \epsilon_{g, 2} \cdot g_t, \forall t$ for some $\epsilon_{g, 2} \geq 0$, $\left| \frac{\sum_{i=-w}^w g_{t+i}}{2w+1} - g_t\right| \leq g_t \cdot E_g(w)$,
    where, $E_g(w) = \frac{w(w+1)}{6}\epsilon_{g, 2} + o(w^4\epsilon_{g, 2}^2)$.
\end{lemma}
This implies that if we apply a moving average smoothing to a time-series $g_t$, the resulting time-series $g_{t, w} = \sum_{i=-w}^w g_{t+i}/(2w+1)$ is close to $g_t$ for small windows. 

\subsubsection{Demonstrating the Advantage of Smoothing}
We will use the fact that typical real-world signals to be estimated, $f_t$, are smooth (Observation~\ref{assume:smooth}) to find better estimates through aggregation/a weighted moving average. Instead of trying to estimate $\mathbb{E}(X_t)$ from data, we can try to estimate some aggregation over a window, $\mathbb{E}(X_{t,w})$. Therefore, we can use more 
responses,  which may decrease sample variance but may also introduce an error as $\mathbb{E}(X_{t,w}) \neq \mathbb{E}(X_{t})$. The following lemma identifies the conditions when such aggregation is better than  individually estimating $\mathbb{E}(X_t)$. We will measure this by finding $\lambda$ so that smoothing ($\bar{X}_{t,w}/\mu_D$) is less likely to result in a \textit{fractional error} (ratio of difference from $f_t$ to $f_t$) greater than some $\lambda$ compared to no smoothing ($\bar{X}_{t}/\mu_D$).

\begin{lemma}\label{lem:extra-dev}
    Let $\bar{X}_{t, w}$ be some linear combination of $\{\bar{X}_{t-w}, \dots, \bar{X}_{t+w}\}$ such that $|\mathbb{E}(\bar{X}_{t, w})/\mu_D - f_t| \leq \lambda'f_t$. Then,
    the probability of fractional error by $\lambda$ is lower in the smoothed response than the unsmoothed response, $P\left(|\frac{\bar{X}_{t, w}}{\mu_D} - f_t| \geq \lambda f_t \right) \leq P\left(|\frac{\bar{X}_{t}}{\mu_D} - f_t| \geq \lambda f_t\right)$
    if 
    $
        \lambda \geq \lambda'/
        \left(1 - \frac{\sigma_{\bar{X}_{t, w}}}{\sigma_{X_t}/\sqrt{n_t}}\right).
    $
\end{lemma}

Lemma~\ref{lem:extra-dev} suggests that \textit{\textbf{an aggregation across the window is good if}} (i) $\lambda'$ is small, i.e.,  $\mathbb{E}(\bar{X}_{t,w})$ does not deviate too much from $\mathbb{E}(X_{t})$, and 
(ii) $\sigma_{\bar{X}_{t, w}} \ll \sigma_{X_t}/\sqrt{n_t}$.

Let $\bar{X}_{t, w}$ be the random variable defined as $\bar{X}_{t, w} = \sum_{i=-w}^w \frac{n_{t+i}}{n_w} \bar{X}_{t+i}$,
where $n_t$ is the number of responses at time $t$ and $n_{w} = \sum_{i=-w}^{w} n_{i}$. To see why this is a good aggregation, we make the following observation.

\begin{observation}\label{assume:resp_ind}
    Over a selected window $w$, the responses at different $t$ are independent of each other.
\end{observation}
At each time $t$ we randomly select individuals to respond to the survey, and therefore the responses are independent. This will be violated in some extreme cases, such as surveying a highly infectious disease where the respondents happen to be the same every day. Then the response from the same person on consecutive surveys may become dependent. However, we assume that such extreme cases do not occur. Further, this provides another guideline for designing such surveys, i.e., avoiding asking a fixed set of individuals.


\begin{assumption}\label{assume:phi}
 Over a selected window $w$, for a pair of nodes $v_1$ and $v_2$, with a common neighbor, $\mathbb{E}(I_{v_1}I_{v_2}) = \phi_t f_t, \forall t$, for some smooth $\phi_t \in [0, 1]$, such that $|\Delta^2 \phi_t| \leq \epsilon_\phi$
\end{assumption}
This is reasonable because the covariance of the infection state of two randomly selected nodes with a common neighbor, should not vary rapidly over time.
Recall that
$
    \sigma_{X_t}^2 =  f_t (\mu_D^2 (\phi_t-f_t) + \mu_D (1-\phi_t) + \sigma_D^2 \phi_t).
$    
 Since all terms of $\sigma_{X_t}^2$ are product of smooth functions, for some $\epsilon_{\sigma^2}, |\Delta \sigma_{X_t}^2| \leq \epsilon_{\sigma^2}$. This is also demonstrated in Figure~\ref{fig:ca_diffs}. For the sake of demonstration, we calculated $\sigma_{X_t}$ by setting 
 $\mu_D = 15, \sigma_D = 10, \phi_t = f_t$. 
 With Observation~\ref{assume:resp_ind} and Assumption~\ref{assume:phi}, we are ready to prove the following theorem.

\begin{theorem}\label{thm:accum}
    The probability of fractional error of $\lambda$ is lower in the smoothed response compared to the unsmoothed response, $P\left(\left|\frac{\bar{X}_{t, w}}{\mu_D} - f_t\right| \geq \lambda f_t \right) \leq P\left(\left|\frac{\bar{X}_{t}}{\mu_D} - f_t\right| \geq \lambda f_t\right)$
    if 
    \begin{equation}\label{eqn:thm_accum1}
        \lambda \geq w\epsilon_{f, 1}/ \left(1 - \left(1+\frac{w\epsilon_{\sigma^2, 1}}{1 - w\epsilon_{\sigma^2, 1}}\right)\sqrt{n_t/n_w}\right).
    \end{equation}
\end{theorem}

The theorem suggests that \textbf{the smoothed response is less likely to deviate} by some small $\lambda$ from the true value compared to the unsmoothed response. The inequality that $\lambda$ needs to satisfy to justify smoothing suggests that if the first differences of the hidden time series and its variance are small, we can consider a larger window to smooth the responses. Further, $n_t \ll n_w$ is desirable.

\subsubsection{Stronger Results when Variance of $n_t$ is Small}
Assume that the number of responses per unit time $n_t$ does not vary drastically over a window.
\begin{assumption}
    $\sigma_n/\mu_n  \ll 1.$
\end{assumption}
This may not be true if we aggregate responses for each day since, within a week, weekdays may have different patterns than weekends. 
However, for aggregated weekly observations, $n_t$ may not vary significantly. Suppose $\mu_n = n_w/(2w+1)$ and $\sigma_n$ represent the mean and standard deviation of $\{n_{t-w}, \dots, n_{t+w}\}$, respectively.

\begin{lemma}\label{lem:small_sigma_n}
For any smoothly varying sequence $g_t$ with bounded second difference, if $\sigma_n/\mu_n < 1$, then $\left|\sum_{i=-w}^{w} \frac{n_{t+i}}{n_w} g_{t +i}  - g_t \right| \leq g_t\gamma_g\,,$
for some small $\gamma_g \geq 0$.
\end{lemma}
This leads to a \textbf{better  error bound} from indirect surveys.
\begin{theorem}\label{thm:accum2}
    The probability of fractional error of $\lambda$ is lower in the smoothed response compared to the unsmoothed response, $P\left(\left|\frac{\bar{X}_{t, w}}{\mu_D} - f_t\right| \geq \lambda f_t \right) \leq P\left(\left|\frac{\bar{X}_{t}}{\mu_D} - f_t\right| \geq \lambda f_t\right)$
    if 
    \begin{equation}\label{eqn:thm_accum2}
        \lambda \geq \gamma_f / \left(1 - \sqrt{\frac{n_t}{n_w}(1+\gamma_{\sigma^2})}\right),
    \end{equation}
    where $\gamma_g = E_{g}(w) + \epsilon_{g, 1}\frac{\sigma_n}{\mu_n}\frac{w\epsilon_{g, 1}}{1- w\epsilon_{g, 1}}$ and $E_g(w) \approx \frac{w(w+1)}{6}\epsilon_{g, 2}$.
\end{theorem}

\begin{figure}[!t]
    \centering
    \includegraphics[width=0.9\columnwidth,trim={25 15 35 15},clip]{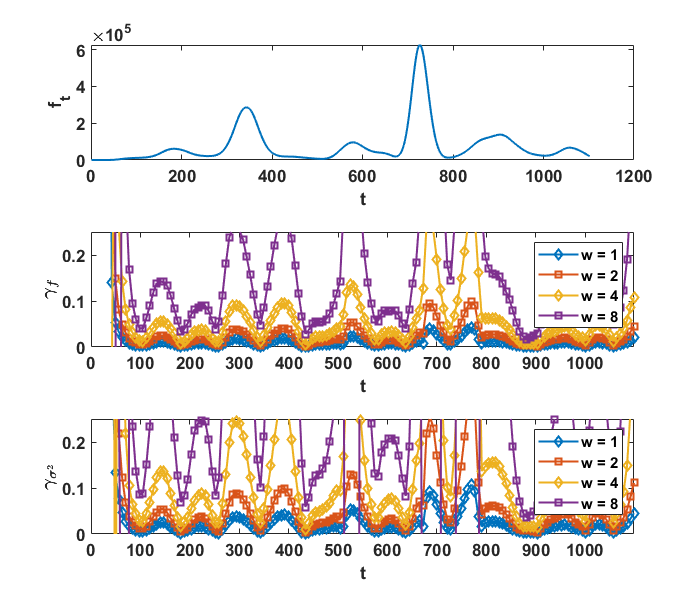}
    \caption{$\gamma_f$ and $\gamma_{\sigma^2}$ calculated over various window sizes for COVID-19 reported cases in California. The values are small as desired, particularly for small $w$.}
    \label{fig:gamma_ca}
\end{figure}

To demonstrate that $\gamma_f$ and $\gamma_{\sigma^2}$ are indeed small, we calculate them for various window sizes over COVID-19 reported cases (Figure~\ref{fig:gamma_ca}).
Their values increase with larger $w$ (recall that the window size is $2w+1$). For these calculations, we set $\sigma_n/\mu_n = 0.3$. For small windows, the values are small, and so smoothing is advantageous. As expected, for large windows,the signal is oversmoothed resulting in higher values of $\gamma_f$ and $\gamma_{\sigma^2}$, consequently higher errors.

\remove{
\subsection{Extension to Biased Reporting}
So far, we have assumed that the response $X_t$ is honest and correct. However, in reality, the reporting could be biased to exaggerate counts or undercount,  or the respondents may incorrectly recall. Suppose, the bias follows the following pattern.
\begin{assumption}\label{assume:bias}
The respondents can be grouped, where an individual belongs to the group $j$ with probability $q_j$ (independently of its in-degree) and overestimates the hidden population by a factor of $\alpha_j$ (underestimates if $\alpha_j$ < 1). Also, $q_j$ and $\alpha_j$ are constants over time.    
\end{assumption}
Suppose $Z_t$ is a randomly selected response with possible bias following Assumption~\ref{assume:bias} and $\bar{Z}_t$ is the sample mean. Then
\begin{theorem}\label{thm:bias}
 $\mathbb{E}(Z_{t}) = B\mu_D f_t$, $\sigma_{Z_t}^2 = B^2\sigma_{X_t}^2$, and for any $\lambda \leq 0$
 \begin{equation}
     P\left(\left|\frac{\bar{Z}_t}{B\mu_D} - f_t \right| > \lambda \right) = P\left(\left|\frac{\bar{X}_t}{\mu_D} - f_t \right| > \lambda \right)\,,
 \end{equation}
where $B = \sum_j q_j \alpha_j$.
\end{theorem}
The proof is available in the Appendix.
 Therefore, all of our results still apply after normalizing the time-series of $\bar{Z}_{t}$ and $\bar{Z}_{t, w}$ even if there are biases in the responses under Assumption~\ref{assume:bias}.

The above analysis does not take into account respondents who are malicious, deliberately inaccurate (possibly due to the sensitivity of the questions), or exhibit other behaviors that deviate from Assumption~\ref{assume:bias}. In practice, additional steps can be taken to handle these different biases~\cite{scheers1992review,kazemzadeh2016frequency,ezoe2012population,salganik2011game}.
}

\section{Results and Analysis}

\subsection{Synthetic Experiments}
\label{sec:simul}

To evaluate our claims and analyze the effect of various variables, we ran an epidemic simulation in conjunction with the simulation of surveys over randomly generated networks\footnote{Our code is available at 
\url{https://github.com/GCGImdea/coronasurveys/tree/master/papers/2024-AAAI-Nowcasting-Temporal-Trends-Using-Indirect-Surveys}.
}.
\paragraph{Epidemic simulation}
We use an extended SIR model to simulate an epidemic with varying infection parameters. The infection parameter starts at a value so that the reproduction number R0~\cite{dietz1993estimation} is above 2. At random times, we introduce ``interventions'' that reduce R0 smoothly to a value below 1. We run several such simulations and pick one that produces multiple peaks over 600 days, to emulate complex realistic epidemics like Influenza and COVID-19 that have multiple waves. We acknowledge that, in reality, not all infections will be detectable. However, assuming that each infection will be detected with a fixed probability only scales the time-series $I(t)$ by a constant. Therefore, for the purpose of this study, we directly use $I(t)$ to compute the hidden population over time.

\paragraph{Survey simulation}
We simulate sampling nodes (respondents) from a graph with a power law distribution $p_k \propto k^{-2}$, with a bounded maximum degree, such that the mean degree is approximately a parameter $d$.  The choice of distribution was driven by the intention to introduce some skewness in the degree distribution to push the limits of our approach. The main conclusions do not change on Erdos-Renyi graphs.
The simulation has the following parameters. (i) $d$: approximate average degree in the latent graph;
(ii) $n$: upper limit on the number of individuals who respond on a given day. The actual number is uniformly selected from $1$ to $n$;
(iii) $n_d$: number of nodes that can potentially be covered by the respondents;
(iv) \textit{accum}: number of days over which the responses are accumulated ;
(v) \textit{period}: time window within which the respondents are to count the hidden population. E.g., if the question is ``how many people do you know who have had COVID-like illnesses in the last 7 days,'' then \textit{period} $= 7$.
For each combination of the above parameters, we ran 16 simulations resulting in \textbf{82,000 combinations} of parameters and simulations. 
For indirect surveys, we randomly infect each neighbor of the responders with the probability $\sum_{\tau=0}^{\mbox{\textit{period}}-1} I(t-\tau)$ and obtain the indirect response from each responding node. We then accumulate the responses obtained over \textit{accum} number of days.  For comparison, we also introduced the traditional NSUM approach \cite{killworth1998social}, where the response from each node is normalized by its degree.
For direct surveys, we infect nodes among the responders and note the number of infections produced.

\subsubsection{Results}
For each of indirect (\textbf{Ind}), NSUM, and direct (\textbf{Dir}) survey methods, we introduced the following post-processing methods:  (1) \textbf{NoS}: Average response for each time unit (defined by \textit{accum}) without any smoothing.
    (2) \textbf{WA}: Moving average of NoS weighted by the number of responses over a window $w$.
    (3) \textbf{UA}: Unweighted moving average of NoS over a window $w$.
These methods were compared against the infections $I(t)$ using MAE with time granularity redefined by the choice of \textit{accum}. Before computing the error, we apply a range normalization, so that the maximum of each time-series is set to 1 and the minimum to 0.  Note that here, $I(t)$ is not the same as $f_t$. To construct $f_t$ we would count the number of infections in the last \textit{period} number of days.
Secondly, note that setting the parameter \textit{accum} $> 1$ is equivalent to performing a weighted average (scaled by a constant factor). Therefore, \textit{accum} $>1$ and $w=0$ will have a similar effect as using $w > 1$ on \textit{accum} $=1$.

\begin{figure}[!t]
    \centering
    \includegraphics[width=0.9\columnwidth, trim={0 2 0 2}, clip]{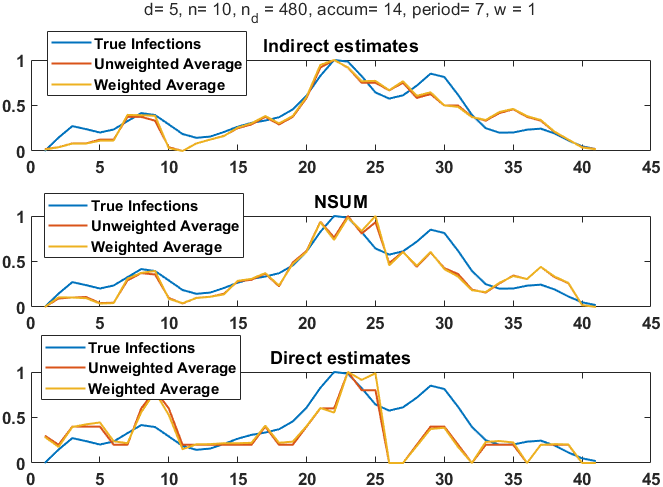}
    \caption{Result of one of the survey simulations.}
    \label{fig:sample-result}
\end{figure}

Figure~\ref{fig:sample-result} shows the result of the survey simulation with parameters $d=5, n=10, n_d=480, accum=14, period=14,$ and $w=15$. Time-series obtained from smoothed indirect survey is much more similar to the true infections $I(t)$ compared to those obtained from direct surveys. The time-series of smoothed NSUM is also close to the true infections, but at times, worse than our indirect method.

\begin{figure}[!t]
    \centering
    \includegraphics[width=\columnwidth,trim={35 0 50 10},clip]{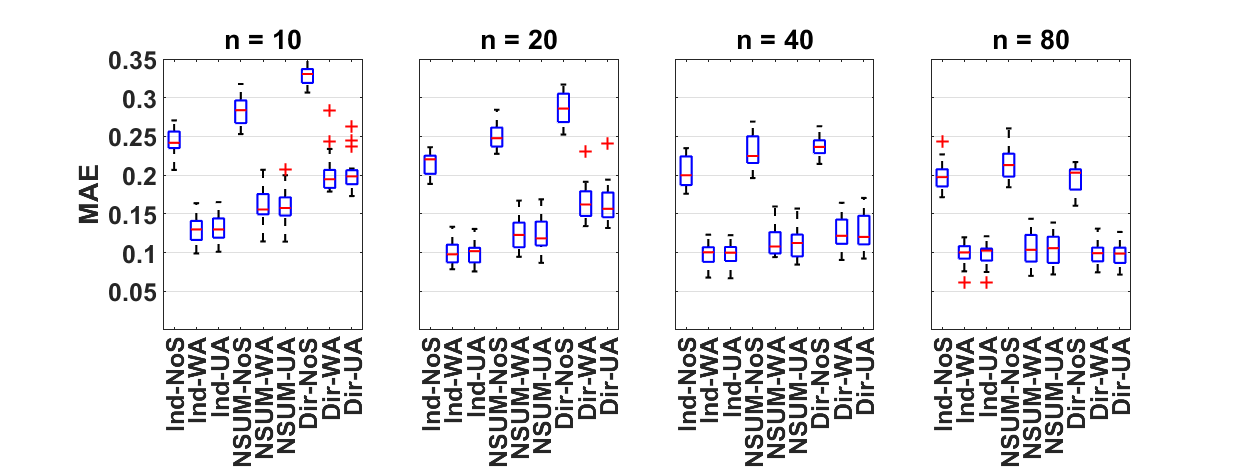}
    \includegraphics[width=\columnwidth,trim={35 0 50 10},clip]{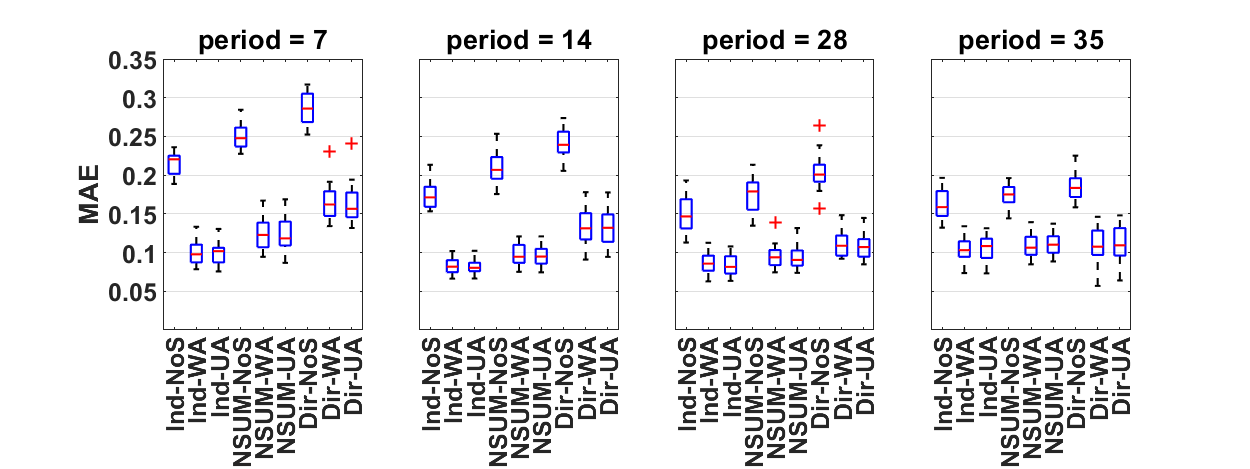}
    \includegraphics[width=\columnwidth,trim={35 0 50 10},clip]{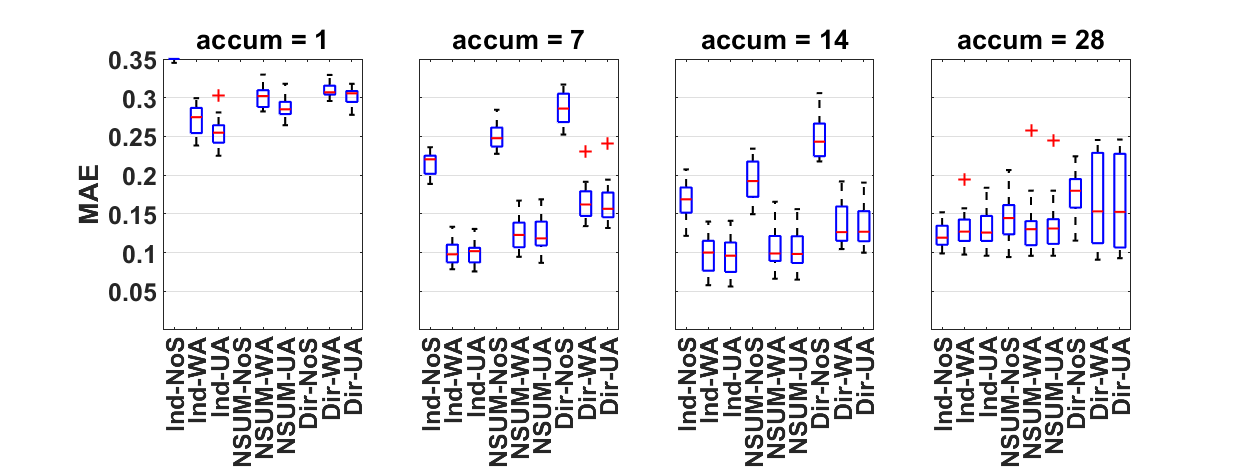}
    \includegraphics[width=\columnwidth,trim={35 0 50 10},clip]{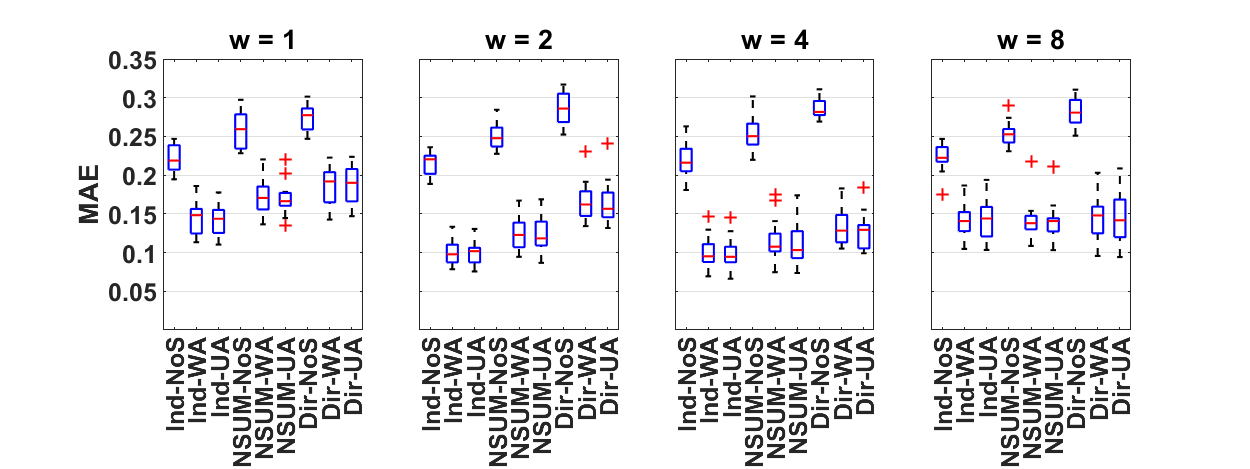}
    \caption{\textbf{1st row:} MAE vs $n$, the number of respondents. Indirect with moving average is the best for low $n$. All moving averages converge as $n$ increases.
    \textbf{2nd row:} MAE vs \textit{period} in the survey question. Smoothed results converge as \textit{period} increases.
    \textbf{3rd row:} MAE vs \textit{accum}, the accumulation width. Errors reduce quickly as \textit{accum} increases. Larger values make the moving averages slightly worse.
    \textbf{4th row:} MAE vs $w$, the smoothing window. Errors reduce as $w$ increases and increase slightly for large $w$.}
    \label{fig:ae_n}
\end{figure}

Figure~\ref{fig:ae_n} shows
the distribution of errors obtained by different methods. In this figure, to focus on the impact of one parameter, we fix the others ($d=5$, $n=20$, $n_d = 60$, $period=7$, $accum=7$, and $w = 2$).
In general, we note that the indirect methods (Ind-*) produce lower median errors than NSUM-* the direct methods (Dir-*). Also, there is no significant difference between weighted and unweighted smoothing strategies. In terms of parameters, the choice of $d$ and $n_d$ do not impact the relative patterns across the 9 methods (see Supplementary Material in full version \cite{srivastava2023estimating}). As expected from our analysis (Theorems~\ref{thm:accum} and~\ref{thm:accum2}), increasing $accum$ first decreases the errors, but a high value worsens the performance for the moving averages (*-UA, *-WA). A similar observation can be made for increasing $w$ -- *-NoS which has $w=0$ produces a higher error than $w=1$. All moving averages become similar as $w$ increases to $8$.

\subsection{Real Dataset}\label{sec:CTIS}
 

The objective of these experiments is to evaluate the performance of the proposed approach using datasets drawn from the US COVID-19 Trends and Impact Survey (CMU-CTIS)~\cite{salomon2021us}.
It has data on self-reported symptoms, symptoms in the respondent's community, testing, isolation measures, vaccination acceptance, and mental health, among other factors, to assess the spread of COVID-19. Approximately 40,000 US respondents participated in this survey daily between April 6, 2020, and June 25, 2022. We chose the period of September 8, 2020, through March 1, 2022 because CMU-CTIS included in it a question regarding individual COVID-19 positive test results, which is used to estimate the direct survey results. A null value detection and removal, as well as an outlier filter 
are applied to the dataset for each state.

\begin{figure}[!t]
    \includegraphics[width=\columnwidth, trim={0 48 0 20}, clip]{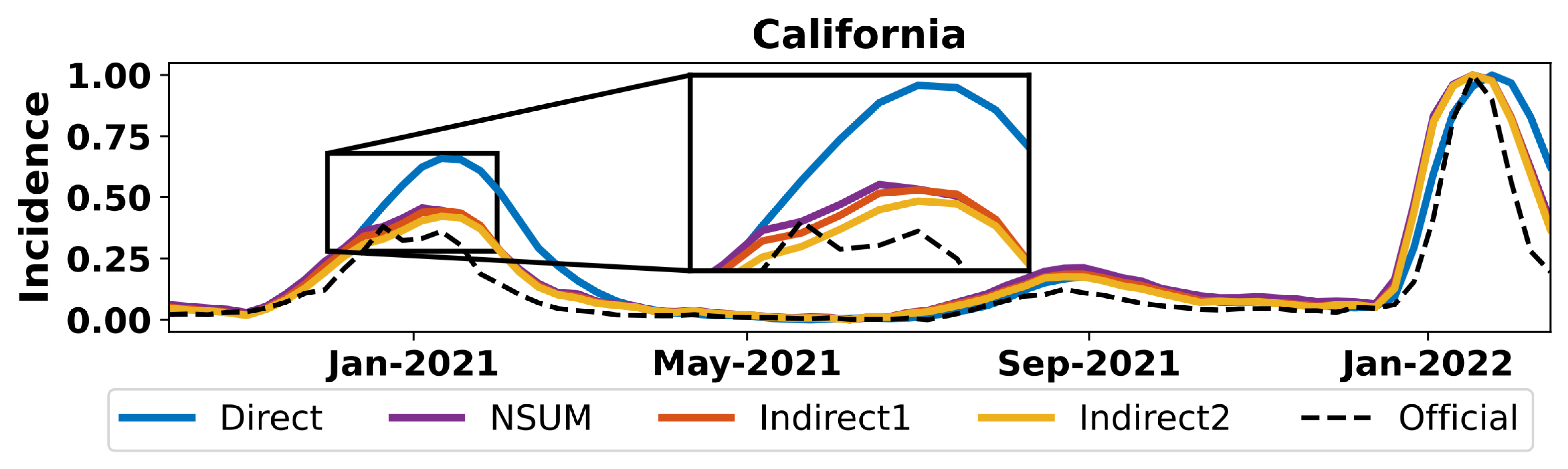} 
\caption{Normalized COVID-19 incidence estimated from the US CMU-CTIS data for California. The parameters of the proposed approach were set to $accum = 7$ and $w=1$.}\label{fig: CTIScrvs}
\end{figure}

Figure \ref{fig: CTIScrvs} shows the normalized COVID-19 incidence curves from direct and indirect surveys for California from September 2020 to March 2022. For the curves derived from indirect surveys, we included both the CLI incidence reported in the household (\textbf{Indirect1}) and the CLI incidence reported within the local community (\textbf{Indirect2}).  In addition, Figure~\ref{fig: CTIScrvs} displays the curve obtained by the NSUM method \cite{killworth1998social} using household questions (CTIS has a question asking for the size of the household). For these curves, we set the parameters to $accum = 7$ and $w = 1$. For comparison purposes, we include the normalized incidence curves obtained from datasets provided by the Johns Hopkins Coronavirus Resource Center \cite{dong2020interactive}. We observe that the curves obtained from indirect surveys are much similar to the official ones. To quantitatively evaluate the proposed approach, Table \ref{tab:maeCTIS} shows the MAE of the normalized incidence curves obtained from direct, NSUM and indirect surveys conducted in California, Texas, New York, and Pennsylvania for a variety of values of $accum$ and $w$. The reference curves are based on official data. For each ($accum$, $w$) pair, a bold font, and underlined values correspond to the best and the second-best values, respectively. As given in Table \ref{tab:maeCTIS}, the incidence curves obtained from indirect surveys exhibit lower MAE values than those obtained by both NSUM and direct approaches. Furthermore, the MAE values obtained from indirect surveys in the local community (Indirect2) are generally lower than those extracted from indirect surveys in the household (Indirect1).

\begin{table}
\caption{MAE of the normalized COVID-19 incidence curves estimated from the US CMU-CTIS data for California, Texas, New York, and Pennsylvania and for different values of $accum$ and $w$.}\label{tab:maeCTIS}
\small
\begin{tabular}{|c| c | c |c c c c |}
\hline
$accum$ & $w$ & state & Direct & NSUM 
& Indirect1 & Indirect2 \\
\hline
\multirow{12}{*}{7} & \multirow{4}{*}{1} & CA & 0.0703 & 0.0443 & \underline{0.0341} & \textbf{0.0292} \\
 & & TX & 0.0661 & 0.0379 & \underline{0.0289} & \textbf{0.0270} \\
 & & NY & 0.0785 & 0.0315 &\underline{0.0301} & \textbf{0.0299} \\
 & & PN & 0.0572 & 0.0368 &\underline{0.0300} & \textbf{0.0263} \\ \cline{2-7}
 & \multirow{4}{*}{3} & CA & 0.1148 & 0.0988 & \underline{0.0881} & \textbf{0.0811} \\
 & & TX & 0.1236 & 0.0890 &\underline{0.0813} & \textbf{0.0782} \\
 & & NY & 0.1210 & 0.0930 & \underline{0.0910} & \textbf{0.0886} \\
 & & PN & 0.0956 & 0.0907 & \underline{0.0816} & \textbf{0.0691} \\
 \hline
 \multirow{12}{*}{14} & \multirow{4}{*}{1} & CA & 0.0836 & 0.0624 & \underline{0.0524} & \textbf{0.0477}  \\
 & & TX & 0.0779 & 0.0385 & \underline{0.0343} & \textbf{0.0336} \\
 & & NY & 0.0929 & 0.0520 & \underline{0.0504} & \textbf{0.0500} \\
 & & PN & 0.0689 & \textbf{0.0389} & \underline{0.0391} & 0.0429 \\ \cline{2-7}
  & \multirow{4}{*}{3} & CA & 0.1441 & 0.1217 & \underline{0.1116} & \textbf{0.1059} \\
 & & TX & 0.1349 & 0.1058 & \underline{0.1042} & \textbf{0.1027} \\
 & & NY & 0.1571 & 0.1165 & \underline{0.1126} & \textbf{0.1090} \\
 & & PN & 0.1349 & 0.1182 & \underline{0.1110} & \textbf{0.1005} \\
\hline
\end{tabular}\vspace{-10pt}
\end{table}


\section{Discussion}
\label{sec:discussion}

\paragraph{Survey Design.} Our assumptions and results can be used to design better indirect surveys. As lower variance in the degree distribution is desirable due to Theorem~\ref{thm:indirect_better}, the question can be framed in a way to keep the variance low. For instance, instead of simply asking ``how many people do you know who $\dots$'', we could restrict the set of people to be counted -- ``Among your and your two immediate neighboring households, how many do you know who $\dots$.''

\paragraph{Targeted Surveys.} Our approach allows the responses to be from a restricted subset of the population. This is advantageous for targeted surveys. For instance, healthcare workers are more likely to know of Influenza hospitalizations than the general public. Therefore, we can restrict the survey to them to estimate the number of hospitalizations over time. Further, any direct survey on an online platform can only estimate the hidden subpopulation among those using the platform. Instead, an indirect survey will cover all the neighbors of the platform users in the latent graph.


\paragraph{Extension to Biased Reporting}
In this paper we have assumed that the survey responses are honest and correct. However, in reality, the reporting could be biased to exaggerate counts or undercount,  or the respondents may incorrectly recall. A detailed analysis of bias is beyond the scope of this paper. However, it can be shown that all our results apply if the respondents have different biases that underestimate or overestimate at the same rate over time -- these biases scale the estimate by a constant factor.
These results do not take into account respondents who are malicious, deliberately inaccurate (possibly due to the sensitivity of the questions), or exhibit other behaviors. In practice, additional steps can be taken to handle different biases~\cite{scheers1992review,kazemzadeh2016frequency,ezoe2012population,salganik2011game}.

\section{Conclusions}
We have proposed a latent graph formulation to estimate the temporal trends in the size of a hidden population from indirect surveys, leading to better estimates than those achievable with direct surveys having the same number of responses. 
We leveraged the temporal dynamics of the underlying process and identified the conditions under which a weighted moving average of responses leads to better estimates compared to raw responses. We performed extensive simulations of a temporal process over which a simulated survey is performed, to study the impact of various parameters on the estimation error. We demonstrated that our approach outperforms traditional Network Scale-Up Methods and the direct approach with and without performing a moving average.
We also demonstrated that our approach is able to better estimate the trend of COVID-19 cases on real-world surveys over time.

\section*{Acknowledgements}

This work was partially supported by grants TED2021-131264B-I00 (SocialProbing) and PID2019-104901RB-I00, funded by MCIN/AEI/10.13039/501100011033 and the European Union “NextGenerationEU”/PRTR.
The authors would like to thank Mohamed Kacem for his contribution to pre-processing the data used in this work.

\section*{Ethical Declaration}

The Ethics Board (IRB) of IMDEA Networks Institute approved this work on 2021/07/05. IMDEA Networks has signed Data Use Agreements with Facebook and Carnegie Mellon University (CMU) to access their data. Specifically, CMU project STUDY2020\_00000162 entitled ILI Community-Surveillance Study. Informed consent has been obtained from all participants in this survey by this institution. All the methods in this study have been carried out in accordance with relevant ethics and privacy guidelines and regulations.

\section*{Availability of Data and Materials}

The data presented in this paper (in aggregated form) and the codes used to process it will be made publicly available at 
\url{https://github.com/GCGImdea/coronasurveys/tree/master/papers/2024-AAAI-Nowcasting-Temporal-Trends-Using-Indirect-Surveys}. 
The microdata of the CTIS survey from which the aggregated data was obtained cannot be shared, as per the Data Use Agreements signed with Facebook and Carnegie Mellon University (CMU).


\bibliography{sample-base,refs}


\clearpage

\appendix
\section*{Supplementary Material}

\section{Proofs}\label{sec:proofs}
Here we provide detailed proofs of all the Theorems and Lemmas used in the paper.
\begin{proof}[Proof of Theorem~\ref{thm:prop}]
Let $p_k$ be the fraction of nodes with in-degree $k$ in $G$. Consider a node partitioning graph $G$ into sets $\{A_1, A_2, \dots\}$, where $A_k$ represents the set of all nodes of in-degree $k$. By Assumption~\ref{assume:indep}, for any $v$ belonging to the neighborhood of some $u \in A_k$, the indicator random variable $I_{vk}$ of the event $v \in H_t$ is independent of $k$, i.e., $\mathbb{E}(I_{vk}) = f_t$. Now, if we pick a random node $u$ from $A_k$ for the indirect survey, due to linearity of expectation,
\begin{align}
    \mathbb{E}(X_t|A_k) &= \mathbb{E}\left(\sum_{(v, u)\in E}
    I_{vk}|A_k \right) \\
    &= \sum_{(v, u)\in E} \mathbb{E}(I_{vk}|A_k) = kf_t\,.
\end{align}
When we randomly select a node $u$ in $G$, $P(u \in A_k) = p_k$. By the law of total expectation, 
\begin{align}
    \mathbb{E}(X_t) &= \sum_k \mathbb{E}(X_t|A_k)P(A_k) \\
    &= \sum_k kf_t \cdot p_k =  (\sum_k kp_k) f_t = \mu_D f_t\,.
\end{align}
\end{proof}

\begin{proof}[Proof of Lemma~\ref{lem:sigma_X}]
We wish to compute the variance using $\sigma_{X_t}^2 = \mathbb{E}(X_t^2) - \left(\mathbb{E}(X_t)\right)^2$. We already have $(\mathbb{E}(X_t))^2 = \mu_D^2f_t^2$. By the law of total expectation, $\mathbb{E}(X_t^2) = \mathbb{E}(X_t^2|A_k)P(A_k)$.

\begin{align}
\mathbb{E}(X_t^2) &= \sum_k \mathbb{E}(X_t^2|A_k)P(A_k) \\
&= \sum_k \mathbb{E}\left(\left(\sum_{(v_i, u), (v_j, u) \in E} I_{v_i}I_{v_j})\right)^2|A_k\right) p_k \\
& = \sum_k \left( \sum_{i=1}^k \mathbb{E}(I_{v_{ik}}^2) + \sum_{\underset{i\neq j}{i,j=1}}^k \mathbb{E}(I_{v_{i}}I_{v_{j}}) \right)p_k  \nonumber\\
& = \sum_k \left( k f_t + \sum_{\underset{i\neq j}{i,j=1}}^k \mathbb{E}(I_{v_{i}} I_{v_{j}}) \right) p_k \label{eq-expected}
\end{align}


Since, $I_{v_{i}}$ and $I_{v_{j}}$ are non-negative, and $I_{v_{i}} I_{v_{j}} \leq \min\{I_{v_{i}}, I_{v_{j}}\}$, 
\begin{equation}\label{eqn:corr_bound}
0 \leq \mathbb{E}(I_{v_{i}}I_{v_{j}}) \leq f_t\,.    
\end{equation}

Putting $\mathbb{E}(I_{v_{i}}I_{v_{j}}) = \phi_t f_t$ in Equation~\ref{eq-expected}, we get
\begin{align*}
    \mathbb{E}(X_t^2) &= \sum_k p_k \left(kf_t + (k^2 - k)\phi_t f_t\right) \\
    &= \mu_D f_t + (\sigma_D^2 + \mu_D^2)\phi_t f_t - \mu_D \phi_t f_t\\
    &= f_t(\mu_D + \sigma_D^2\phi_t + \mu_D^2\phi_t - \mu_D \phi_t)\,.
\end{align*}
Therefore,
\begin{align}
    \sigma_{X_t}^2 &= \mathbb{E}(X_t^2) - \left(\mathbb{E}(X_t)\right)^2\nonumber \\
    &= f_t(\mu_D + \sigma_D^2\phi_t + \mu_D^2\phi_t - \mu_D \phi_t) - \mu_D^2 f_t^2 \nonumber \\
    &=  f_t (\mu_D^2 (\phi_t-f_t) + \mu_D (1-\phi_t) + \sigma_D^2 \phi_t)\,.
\end{align}
Since, $0 \leq \phi_t \leq 1$, we have
$ \mu_D f_t (1-\mu_D f_t) \leq \sigma_{X_t}^2 \leq f_t ( \sigma_D^2 + \mu_D^2 (1-f_t)) $.
\end{proof}

\begin{proof}[Proof of Theorem~\ref{thm:indirect_better}]
By Lemma~\ref{assume:clt},
    \begin{align}
        P(|\bar{Y}_t - f_t| > \lambda) 
        &= P\left(\left|\frac{\bar{Y}_t - f_t}{\sigma_{Y_t}/\sqrt{n}}\right| > \frac{\sqrt{n}\lambda}{\sigma_{Y_t}}\right)
        = 1 - erf\left(\frac{\sqrt{n}\lambda}{\sigma_{Y_t}}\right)\,, \label{eqn:direct_dev}
    \end{align}
where $erf$ is the error function, and the equation follows from the fact that $\left(\frac{\bar{Y}_t - f_t}{\sigma_{Y_t}/\sqrt{n}}\right)$ follows standard normal distribution. Similarly,
\begin{align}
        P(|\bar{X}_t/\mu_D - f_t| > \lambda) 
        &= P\left(\left|\frac{\bar{X}_t - \mu_D f_t}{\sigma_{X_t}/\sqrt{n}}\right| > \frac{\mu_D\sqrt{n}\lambda}{\sigma_{X_t}}\right) \nonumber \\
        &= 1 - erf\left(\frac{\mu_D \sqrt{n}\lambda}{\sigma_{X_t}}\right)\,, \label{eqn:indirect_dev}
    \end{align}
Suppose, the probability of deviation is higher from the indirect estimate. Then, from Equations~\ref{eqn:direct_dev} and~\ref{eqn:indirect_dev}, 
\begin{align*}
    & P(|\bar{X}_t/\mu_D - f_t| > \lambda) > P(|\bar{Y}_t - f_t| > \lambda) \\
    & \implies 1 - erf\left(\frac{\mu_D \sqrt{n}\lambda}{\sigma_{X_t}}\right) > 1 - erf\left(\frac{\sqrt{n}\lambda}{\sigma_{Y_t}}\right) \\ 
    & \implies erf\left(\frac{\mu_D \sqrt{n}\lambda}{\sigma_{X_t}}\right) < erf\left(\frac{\sqrt{n}\lambda}{\sigma_{Y_t}}\right) \\ 
\end{align*}
Since, $erf$ is an increasing function,
\begin{equation}\label{eqn:sigma_compare}
    \frac{\mu_D \sqrt{n}\lambda}{\sigma_{X_t}} < \frac{\sqrt{n}\lambda}{\sigma_{Y_t}} 
    \implies \sigma_{X_t} > \mu_D \sigma_{Y_t} \implies \sigma_{X_t}^2 > \mu_D^2 \sigma_{Y_t}^2\,.
\end{equation}
Putting the value of $\sigma_{X_t}^2$ from Lemma~\ref{lem:sigma_X}, combining with Equation~\ref{eqn:sigma_compare}, after some algebraic manipulations gives
\begin{equation}
    \sigma_D^2 > \mu_D (\mu_D - 1)(1-\phi_t)/\phi_t.
\end{equation}
Hence, if $\sigma_D^2 \leq \mu_D (\mu_D - 1)(1-\phi_t)/\phi_t$, then $P(|\bar{X}_t/\mu_D - f_t| > \lambda) \leq P(|\bar{Y}_t - f_t| > \lambda)$.
\end{proof}

\begin{proof}[Proof of Lemma~\ref{lem:first_diff}]
First, assume that $j>0$.
\begin{align}
    |g_{t+j} &- g_t| = |g_{t+j} - g_{t+(j-1)} + g_{t+(j-1)} + \dots + g_{t+1} - g_t| \nonumber\\
    &\mbox{[Adding and subtracting the terms $g_{t+(j-1)}, \dots, g_{t-1}$]} \nonumber\\
    &\leq  |g_{t+j} - g_{t+(j-1)}| + |g_{t+(j-1)} - g_{t+(j-2)}| + \dots + |g_{t+1} - g_t|  \nonumber\\
    &\leq \epsilon_{g, 1} \sum_{i=1}^{j} g_{t+i}
    \leq \epsilon_{g, 1} j g_{max} \label{eqn:first_diff_j}\,.
\end{align}
Since the index of $g_{max}$ must be less than $j$, we can reapply the above as
\begin{align}
    g_{max} - g_t \leq j\epsilon_{g, 1}g_{max} \implies g_{max} \leq \frac{1}{1-j\epsilon_{g, 1}}g_t \label{eqn:first_diff_max}
\end{align}
From Equations~\ref{eqn:first_diff_j} and~\ref{eqn:first_diff_max}, we get
\begin{equation}
    |g_{t+j} - g_t| \leq \frac{j\epsilon_{g, 1}}{1 - j\epsilon_{g, 1}}gt\,.
\end{equation}
Proof for $j<0$ is similar.
\end{proof}


\begin{proof}[Proof of Lemma~\ref{lem:second_diff}]
    Let $S_j = g_{t+j} + g_{t-j}$. Since, $\Delta^2 g_t \leq \epsilon_{g, 2}g_t, \forall t$, we have
    \begin{align*}
        g_{t+1} - 2g_t + g_{t-1} &\leq \epsilon_{g, 2}g_t \\
        g_{t+1} + g_{t-1} &\leq 2g_t + \epsilon_{g, 2}g_t 
        \implies S_1 \leq 2g_t + \epsilon_{g, 2}g_t\,.
    \end{align*}
    And,
    \begin{align*}
        S_2 &= g_{t+2} + g_{t-2}  \\
        &\leq  2g_{t+1} - g_{t} + \epsilon_{g,2}g_{t+1}    \\
        &+ 2g_{t-1} - g_{t} + \epsilon_{g,2}g_{t-1} = 2S_1 - 2g_t + \epsilon_{g,2} S_1 
    \end{align*}
    More generally,
    \begin{align*}
        S_j &= g_{t+j} + g_{t-j}  \\
        &\leq  2g_{t+(j-1)} - g_{t+(j-2)} + \epsilon_{g,2}g_{t+(j-1)} \\
        &+ 2g_{t-(j-1)} - g_{t-(j-2)} + \epsilon_{g,2}g_{t-(j-1)}\\
        &= 2S_{j-1} - S_{j-2} + \epsilon_{g,2}S_{j-1}\\
\implies S_j - S_{j-1} &\leq S_{j-1} - S_{j-2} + \epsilon_{g,2}S_{j-1}
    \end{align*}
Telescoping the above to $j=2$, we get
\begin{align*}
    S_j - S_1 \leq S_{j-1} - 2g_t + \epsilon_{g, 2} \sum_{i=1}^{j-1} S_i
\end{align*}
Since, $S_1 \leq 2g_t + \epsilon_{g, 2}g_t$, we get
\begin{align*}
    S_j - S_{j-1} & \leq 2g_t + \epsilon_{g, 2}g_t - 2g_t + \epsilon_{g, 2} \sum_{i=1}^{j-1} S_i \\
     S_j - S_{j-1} & \leq \epsilon_{g, 2}g_t + \epsilon_{g, 2} \sum_{i=1}^{j-1} S_i \\
    &\leq \epsilon_{g, 2}g_t + \epsilon_{g, 2} (T_{j-1} - g_t) = \epsilon_{g, 2} T_{j-1}\,,
\end{align*}
Where $T_{j} = \sum_{i=-j}^j g_{t+j}$. 
Telescoping again down to $j=2$, we get
\begin{align}
    S_j &\leq S_ 1 + \epsilon_{g, 2}\sum_{i=1}^{j-1} T_i  \nonumber\\
    \implies T_j &\leq T_{j-1} + (2 + \epsilon_{g, 2})g_t + \epsilon_{g, 2}\sum_{i=1}^{j-1} T_i\, \label{eqn:Tj}
\end{align}
for $j>1$, and $T_1 = g_t + S_1 \leq (3 + \epsilon_{g, t})g_t$

We can prove that 
\begin{equation}
    T_w \leq (2w+1)\left(1 + \frac{w(w+1)}{6}\epsilon_{g, 2} + o(w^4\epsilon_{g, 2}^2)\right)g_t\,.
\end{equation}
We omit the details for brevity. The idea is to use induction with base case $w=1$ for which $T_1 \leq (3+\epsilon_{g, 2})g_t \leq \left((2\cdot 1 + 1) + \epsilon_{g, 2} 1\cdot 2\cdot 3/6\right)g_t$.
Therefore,
\begin{align*}
     &\frac{\sum_{i=-j}^j g_{t+j}}{2w+1} - g_t \\
     &\leq \frac{(2w+1)\left(1 + \frac{w(w+1)}{6}\epsilon_{g, 2} + o(w^4\epsilon_{g, 2}^2)\right)g_t}{2w+1} - g_t \\
     &\leq \left(\frac{w(w+1)}{6}\epsilon_{g, 2} + o(w^4\epsilon_{g, 2}^2)\right)g_t\,.
\end{align*}
Repeating the above analysis replacing $\epsilon_{g,2}$ with $-\epsilon_{g,2}$ and reversing all inequalities, 
\begin{align*}
    \frac{\sum_{i=-j}^j g_{t+j}}{2w+1} - g_t 
     \geq -\left(\frac{w(w+1)}{6}\epsilon_{g, 2} + o(w^4\epsilon_{g, 2}^2)\right)g_t\,.
\end{align*}
which completes the proof.
\end{proof}

\begin{proof}[Proof of Lemma~\ref{lem:extra-dev}]
By using triangle inequality, followed by the bound on $\lambda'$, we have
\begin{align*}
    \left|\frac{\bar{X}_{t, w}}{\mu_D} - f_t\right| 
    &\leq \left|\frac{\bar{X}_{t, w}}{\mu_D} - \frac{\mathbb{E}(\bar{X}_{t, w})}{\mu_D}\right| + \left| \frac{\mathbb{E}(\bar{X}_{t, w})}{\mu_D} - f_t\right| \\
    &\leq \left|\frac{\bar{X}_{t, w}}{\mu_D} - \frac{\mathbb{E}(\bar{X}_{t, w})}{\mu_D}\right| + \lambda'f_t \\
    &\leq \left|\frac{\bar{X}_{t, w}}{\mu_D} - \frac{\mathbb{E}(\bar{X}_{t, w})}{\mu_D}\right| + \lambda\left( 1 - \frac{\sigma_{\bar{X}_{t, w}}}{\sigma_{X_t}/\sqrt{n_t}}\right) f_t\,
\end{align*}
Using this inequality, we get
\begin{align}
    &P\left(\left|\frac{\bar{X}_{t, w}}{\mu_D} - f_t\right|\geq \lambda f_t \right) \nonumber\\
    &\leq P\left(\left|\frac{\bar{X}_{t, w}}{\mu_D} - \frac{\mathbb{E}(\bar{X}_{t, w})}{\mu_D}\right| + \lambda\left( 1 - \frac{\sigma_{\bar{X}_{t, w}}}{\sigma_{X_t}/\sqrt{n_t}}\right) f_t \geq \lambda f_t \right) \nonumber\\
    &\leq P\left(\left|\frac{\bar{X}_{t, w}}{\mu_D} - \frac{\mathbb{E}(\bar{X}_{t, w})}{\mu_D}\right| \geq \frac{\lambda f_t \sigma_{\bar{X}_{t, w}}}{\sigma_{X_t}/\sqrt{n_t}} \right) \nonumber\\
    &\leq P\left( \left| \frac{\bar{X}_{t, w} - \mathbb{E}({X}_{t, w})}{\sigma_{\bar{X}_{t, w}}} \right| \geq \frac{\lambda f_t \mu_D}{\sigma_{X_t}/\sqrt{n_t}} \right)
    = 1 - erf\left( \frac{\lambda f_t \mu_D}{\sigma_{X_t}/\sqrt{n_t}} \right) \label{eqn:lin-comb-norm} \\  
    &= P\left(\left|  \frac{\bar{X_t} - \mu_D f_t}{\sigma_{X_t}/\sqrt{n_t}}\right| \geq \frac{\lambda f_t \mu_D}{\sigma_{X_t}/\sqrt{n_t}} \right)
    = P\left(\left|\bar{X_t}/\mu_D - f_t \right| \geq \lambda f_t \right)  \,.  
\end{align}
Equation~\ref{eqn:lin-comb-norm} follows from the fact that $\bar{X}_{t, w}$, being a linear combination of normally distributed variables, also follows a normal distribution.
\end{proof}



\begin{proof}[Proof of Theorem~\ref{thm:accum}]
By linearity of expectation, 
\begin{align}\label{eqn:exp_Xtw}
    \mathbb{E}(\bar{X}_{t, w}) 
    &= \sum_{i=-w}^w \frac{n_{t+i}}{n_w} \mathbb{E}(X_{t+i})
    = \sum_{i=-w}^w \frac{n_{t+i}}{n_w} \mu_D f_{t+i}\,.
\end{align}
Using Lemma~\ref{lem:first_diff},
\begin{align*}
    \mathbb{E}(\bar{X}_{t, w})/\mu_D - f_t &=  \sum_{i=-w}^w \frac{n_{t+i}}{n_w}  f_{t+i} - f_t \\
    &\leq \max_{i=-w}^w{f_{t+i}} - f_t \leq \frac{w\epsilon_{f,1}}{1 - w\epsilon_{f,1}} f_t\,.
\end{align*}
Similarly, $\mathbb{E}(\bar{X}_{t, w})/\mu_D - f_t \geq -w\epsilon_f f_t$, which gives
\begin{equation}
    \left| \mathbb{E}(\bar{X}_{t, w})/\mu_D - f_t \right| \leq \frac{w\epsilon_{f,1}}{1 - w\epsilon_{f,1}} f_t \implies \lambda' = \frac{w\epsilon_{f,1}}{1 - w\epsilon_{f,1}}\,.
\end{equation}

Now,
\begin{align}
\sigma_{\bar{X}_{t, w}}^2
& = \sum_{i=-w}^w \frac{n_i^2}{n_w^2} \sigma_{\bar{X}_{t+i}}^2
= \sum_{i=-w}^w \frac{n_i^2}{n_w^2} \frac{\sigma_{{X}_{t+i}}^2}{n_i} \\
&= \frac{1}{n_w}\sum_{i=-w}^w  \frac{n_i}{n_w}\sigma_{{X}_{t+i}}^2 \label{eqn:sig_Xtw}\\
&\leq \frac{1}{n_w} \max_{i=-w}^w \sigma_{{X}_{t+i}}^2 
\leq \frac{1+\frac{w\epsilon_{\sigma^2, 1}}{1 - w\epsilon_{\sigma^2, 1}}}{n_w} \sigma_{{X}_{t}}^2 \nonumber \\
\implies & \frac{\sigma_{\bar{X}_{t, w}}}{\sigma_{X_t}/{\sqrt{n_t}}}
\leq \sqrt{\frac{n_t}{n_w}}\left(1+\frac{w\epsilon_{\sigma^2, 1}}{1 - w\epsilon_{\sigma^2, 1}}\right) \nonumber\\
&\mbox{Therefore, }\,\, \lambda \geq \frac{\frac{w\epsilon_{f,1}}{1 - w\epsilon_{f,1}}}{1 - \left(1+\frac{w\epsilon_{\sigma^2, 1}}{1 - w\epsilon_{\sigma^2, 1}}\right)\sqrt{n_t/n_w}}.
\end{align}
\end{proof}


\begin{proof}[Proof of Lemma~\ref{lem:small_sigma_n}]
\begin{align}
&\frac{1}{n_w}\sum_{i=-w}^{w} n_{t+i} g_{t +i}
\leq \frac{1}{n_w}\sum_{i=-w}^{w} (\mu_n + n_t - \mu_n) g_{t+i} \nonumber \\
&\leq \frac{1}{\mu_n(2w+1)}\sum_{i=-w}^{w} \mu_n g_{t+i} + \frac{1}{\mu_n(2w+1)}\sum_{i=-w}^{w} (n_{t+i} - \mu_n) g_{t+i} \nonumber\\
&\leq g_{t}(1+\epsilon_{g, 2}) + \frac{1}{\mu_n(2w+1)}\sum_{i=-w}^{w} (n_{t+i} - \mu_n) g_{t+i} \nonumber\\
&\mbox{[By Lemma~\ref{lem:second_diff}]}\nonumber \\
&\leq g_{t}(1+E_{g}(w)) + \frac{1}{\mu_n(2w+1)}\sum_{i=-w}^{w} (n_{t+i} - \mu_n) (g_{t+i} - g_t + g_t) \nonumber\\
&\leq g_{t}(1+E_{g}(w)) + \frac{1}{\mu_n(2w+1)}\sum_{i=-w}^{w} (n_{t+i} - \mu_n) (g_{t+i} - g_t ) \nonumber\\
&\mbox{[Since $g_t\sum_{i=-w}^w (n_{t+i} - \mu_n) = 0$]} \nonumber\\
&\leq g_{t}(1+E_{g}(w))+ \sqrt{\sum_{i=-w}^{w} \frac{1}{\mu_n^2}\frac{(n_{t+i} - \mu_n)^2}{{2w+1}}\sum_{i=-w}^{w} \frac{(g_{t+i}-g_t)^2}{{2w+1}}}\,\nonumber \\
&\hspace{3cm}\mbox{[By Cauchy-Schwarz Inequality]} \nonumber\\
&\leq g_{t}(1+E_{g}(w)) \\
&+ \sqrt{\sum_{i=-w}^{w} \frac{1}{\mu_n^2}\frac{(n_{t+i} - \mu_n)^2}{2w+1}}\max_{i=-w}^w (g_{t+i} - g_i)\,\nonumber \\
&\leq g_{t}(1+E_{g}(w)) + \frac{\sigma_n}{\mu_n}g_{t}\frac{w\epsilon_{g, 1}}{1- w\epsilon_{g, 1}}
\hspace{1cm}\mbox{[By Lemma~\ref{lem:first_diff}]} \nonumber\\
&\leq g_{t}\left(1 + E_{g}(w) + \epsilon_{g, 1}\frac{\sigma_n}{\mu_n}\frac{i\epsilon_{g, 1}}{1- i\epsilon_{g, 1}}\right).
\end{align}
Setting $\gamma_g = E_{g}(w) + \epsilon_{g, 1}\frac{\sigma_n}{\mu_n}\frac{w\epsilon_{g, 1}}{1- w\epsilon_{g, 1}}$,
\begin{equation}
    \frac{1}{n_w}\sum_{i=-w}^{w} n_{t+i} g_{t +i} - g_t \leq \gamma_g g_t
\end{equation}
Similarly, we can show 
\begin{equation}
    \frac{1}{n_w}\sum_{i=-w}^{w} n_{t+i} g_{t +i} - g_t \geq \gamma_g g_t
\end{equation}
\end{proof}

\begin{proof}[Proof of Theorem~\ref{thm:accum2}]
Using Lemma~\ref{lem:small_sigma_n} in Equation~\ref{eqn:exp_Xtw} and~\ref{eqn:sig_Xtw}, we get
\begin{equation}
    \mathbb{E}\left(\left|\bar{X}_{t, w}/\mu_D - f_t\right|\right) \leq \gamma_f f_t\,,
\mbox{ and }
    \left| \sigma^2_{\bar{X}_{t, w}} - \sigma^2_{X_t}\right| \leq \gamma_{\sigma^2} \sigma_{X_t}^2\,.
\end{equation}
Putting these values in Lemma~\ref{lem:extra-dev}, we must have
\begin{equation*}
\lambda \geq \frac{\gamma_f}{1 - \sqrt{\frac{n_t}{n_w}(1+\gamma_{\sigma^2})}}
\end{equation*}
\end{proof}

\subsection{Extension to Biased Reporting}
In this paper, we have assumed that the response $X_t$ is honest and correct. However, in reality, the reporting could be biased to exaggerate counts or undercount,  or the respondents may incorrectly recall. Suppose, the bias follows the following pattern.

\begin{assumption}\label{assume:bias}
The respondents can be grouped, where an individual belongs to the group $j$ with probability $q_j$ (independently of its in-degree) and overestimates the hidden population by a factor of $\alpha_j$ (underestimates if $\alpha_j$ < 1). Also, $q_j$ and $\alpha_j$ are constants over time.    
\end{assumption}
Suppose $Z_t$ is a randomly selected response with possible bias following Assumption~\ref{assume:bias} and $\bar{Z}_t$ is the sample mean. Then
\begin{theorem}\label{thm:bias}
 $\mathbb{E}(Z_{t}) = B\mu_D f_t$, $\sigma_{Z_t}^2 = B^2\sigma_{X_t}^2$, and for any $\lambda \leq 0$
 \begin{equation}
     P\left(\left|\frac{\bar{Z}_t}{B\mu_D} - f_t \right| > \lambda \right) = P\left(\left|\frac{\bar{X}_t}{\mu_D} - f_t \right| > \lambda \right)\,,
 \end{equation}
where $B = \sum_j q_j \alpha_j$.
\end{theorem}

\begin{proof}
The expectation of response $Z_t$ from a randomly selected person is given by
\begin{align}
\mathbb{E}(Z_t) & = \sum_k \mathbb{P}(\delta(u)=k) \sum_j q_j \alpha_j \mathbb{E}\left( \sum_{v \in in(u)} I_v | \delta(u)=k \right) \nonumber\\
& = \sum_k \mathbb{P}(\delta(u)=k) \sum_j q_j \alpha_j \sum_{i=1}^k \mathbb{E}(  I_{v_{ik}} | \delta(u)=k)  \nonumber\\
& = \sum_k p_k \sum_j q_j \alpha_j \sum_{i=1}^k \mathbb{E}( I_{v_{ik}}) \label{eqn:degree-dep2}  \nonumber\\
&= (\sum_j q_j \alpha_j)\sum_k k p_k f_t = f_t B \mu_D\,
\end{align}
where $B = \sum_j q_j \alpha_j$. This suggests that consistent biases among the groups scales the time-series of expected responses by a constant
Similarly, we can show that $\sigma_{Z_t}^2 = B^2 \sigma_{X_t}^2$. Considering the probability of a certain deviation obtained by scaling of the sample mean $\bar{Z}_t$:

\begin{eqnarray}
P\left(\left|\frac{\bar{Z}_t}{B\mu_D} - f_t \right| > \lambda \right) 
        \hspace{-8pt}& = &
        P\left(\left|\frac{\bar{X}_t - \mu_D B f_t}{\sigma_{Z_t}/\sqrt{n}}\right| > \frac{\mu_D\sqrt{n}\lambda}{\sigma_{X_t}}\right) \nonumber \\
        &= & 1 - erf\left(\frac{\mu_D B \sqrt{n}\lambda}{B\sigma_{X_t}}\right)  \nonumber \\
        &= & 1 - erf\left(\frac{\mu_D \sqrt{n}\lambda}{\sigma_{X_t}}\right)\, \label{eqn:indirect_bias_dev}
\end{eqnarray}
which is identical to the expression in Equation~\ref{eqn:indirect_dev} where bias is absent. 
\end{proof}

 Therefore, all of our results still apply after normalizing the time-series of $\bar{Z}_{t}$ and $\bar{Z}_{t, w}$ even if there are biases in the responses under Assumption~\ref{assume:bias}.


The above analysis does not take into account respondents who are malicious, deliberately inaccurate (possibly due to the sensitivity of the questions), or exhibit other behaviors that deviate from Assumption~\ref{assume:bias}. In practice, additional steps can be taken to handle these different biases~\cite{scheers1992review,kazemzadeh2016frequency,ezoe2012population,salganik2011game}.

\section{Additional Results}

\begin{figure}[!t]
    \includegraphics[width=\linewidth]{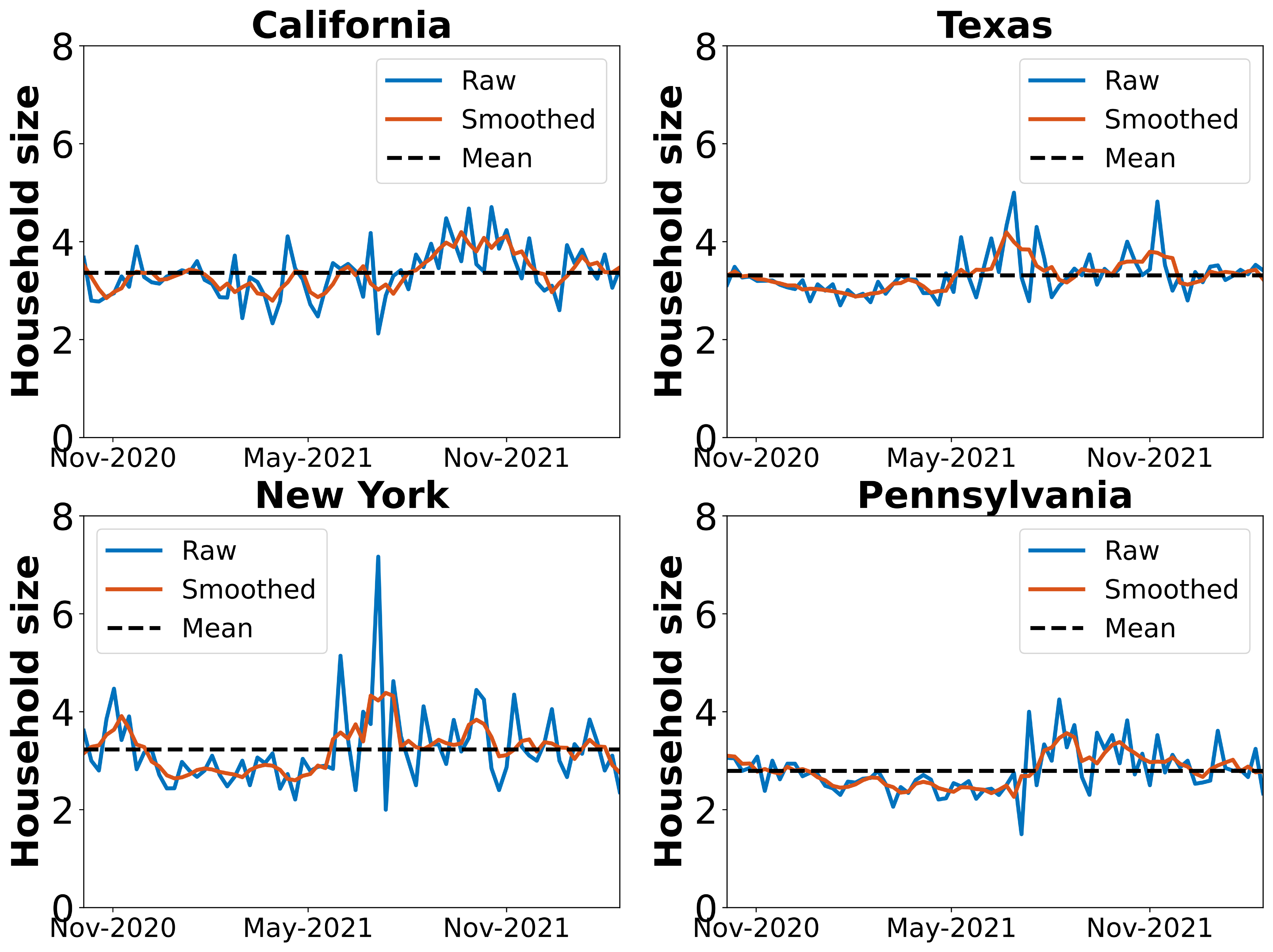}
    \caption{Household size obtained from CMU-CTIS data for California, Texas, New York, and Pennsylvania between September 2020 and March 2022.}
    \label{fig:mu_D}
\end{figure}

\paragraph{Testing Observation~\ref{assume:mu_const}} In our analysis, we indicated that the average degree of the latent graph remains roughly constant. The CMU-CTIS recorded indirect survey information on COVID-like illness (CLI) incidence both in the household and in the local community. Figure~\ref{fig:mu_D} shows the average household size obtained across time from four states (California, Texas, New York, and Pennsylvania). The average remains roughly constant, supporting Observation~\ref{assume:mu_const}.

\begin{figure}[!t]
\begin{tabular}{c}
    \hspace{-10pt}
    \includegraphics[width=\linewidth]{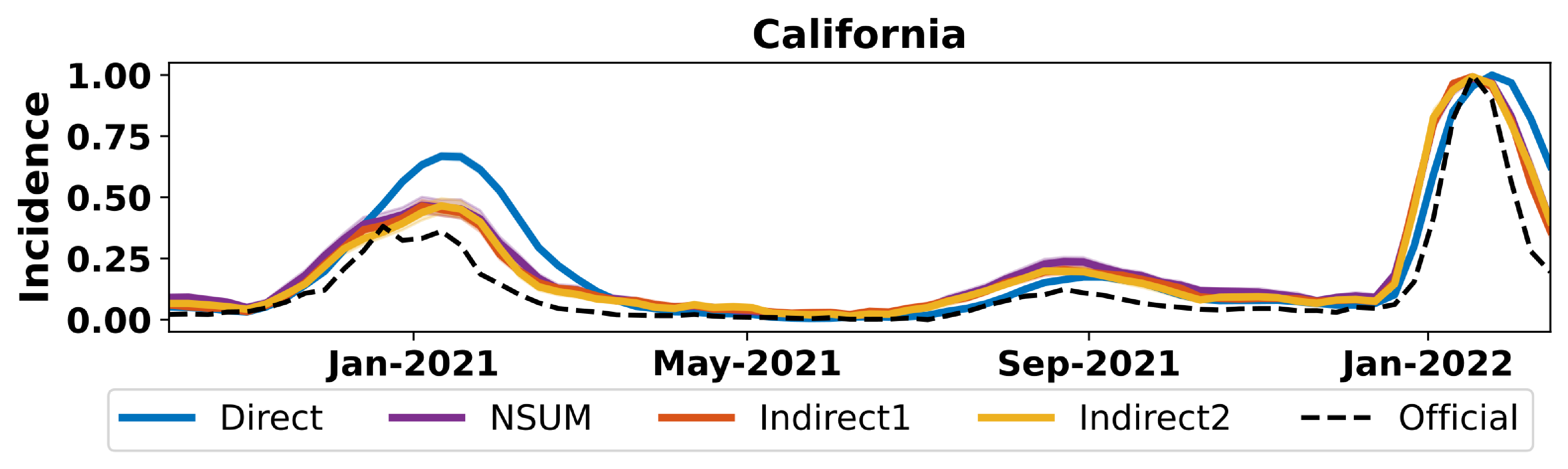} 
    \\
    \hspace{-10pt}
    \includegraphics[width=\linewidth]{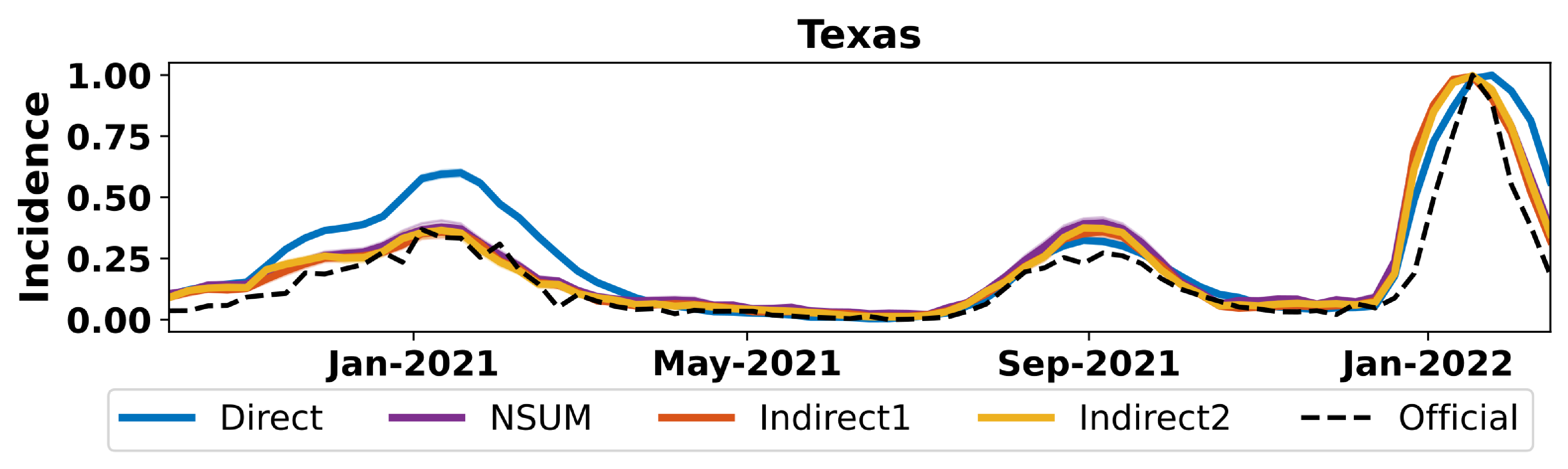} 
    \\
    \hspace{-10pt}
    \includegraphics[width=\linewidth]{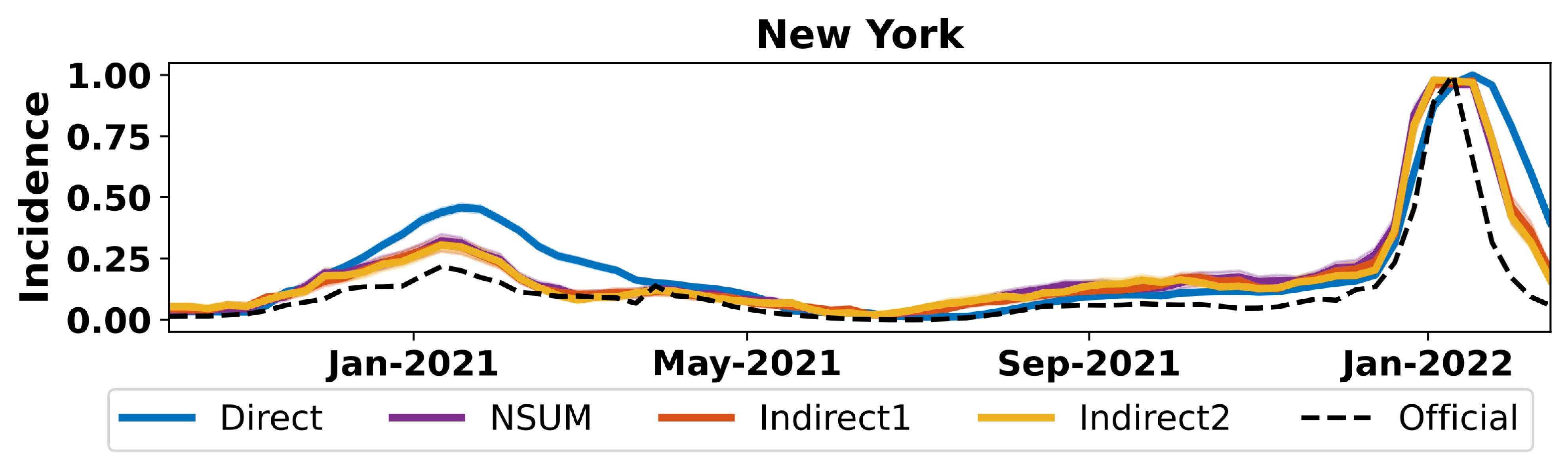} 
    \\
    \hspace{-10pt}
    \includegraphics[width=\linewidth]{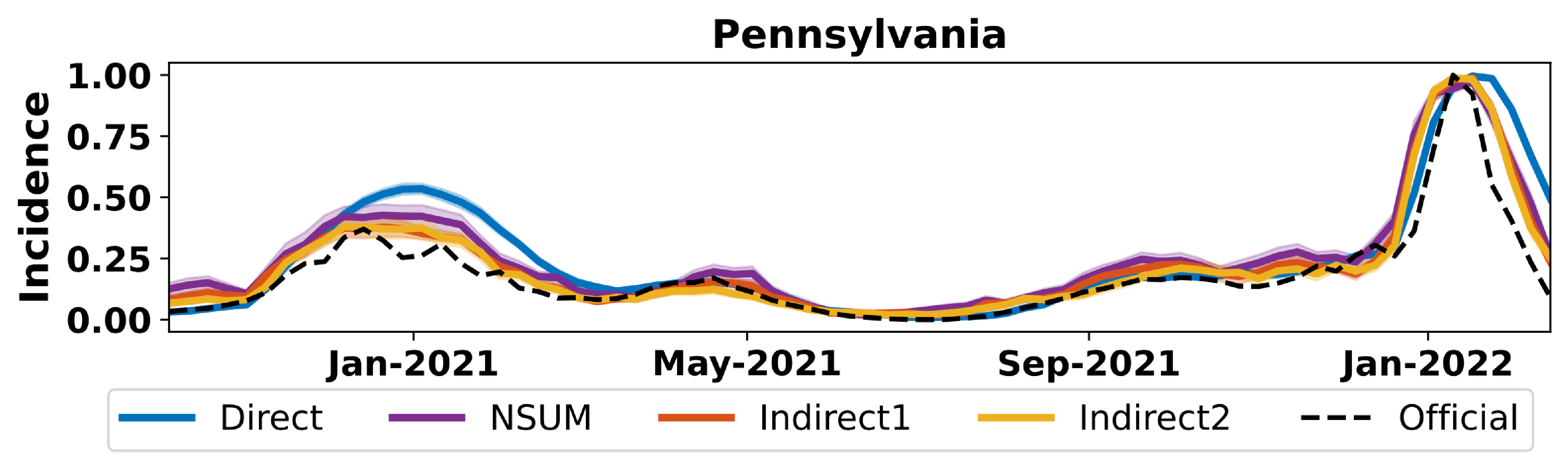}
\end{tabular}
\caption{Normalized COVID-19 incidence curves and the 95\% confidence interval using $10\%$ of the daily samples for the four states and for $accum = 7$ and $w=2$.}\label{fig: CTISsr10}
\end{figure}

\begin{figure}[!t]
\begin{tabular}{c}
    \hspace{-10pt}
    \includegraphics[width=\linewidth]{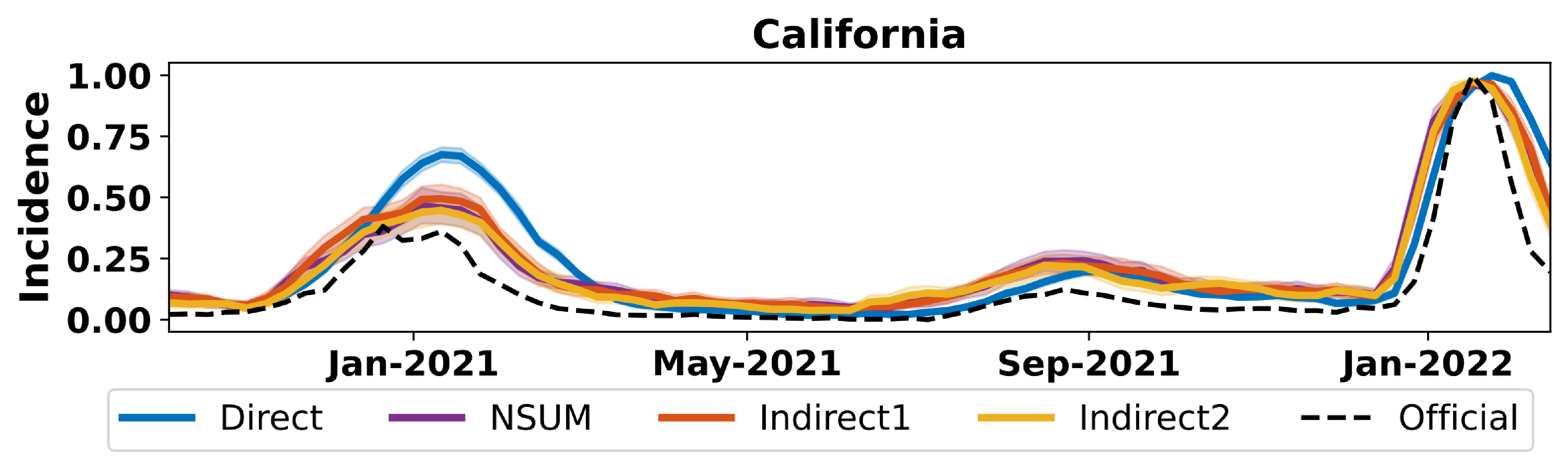} 
    \\
    \hspace{-10pt}
    \includegraphics[width=\linewidth]{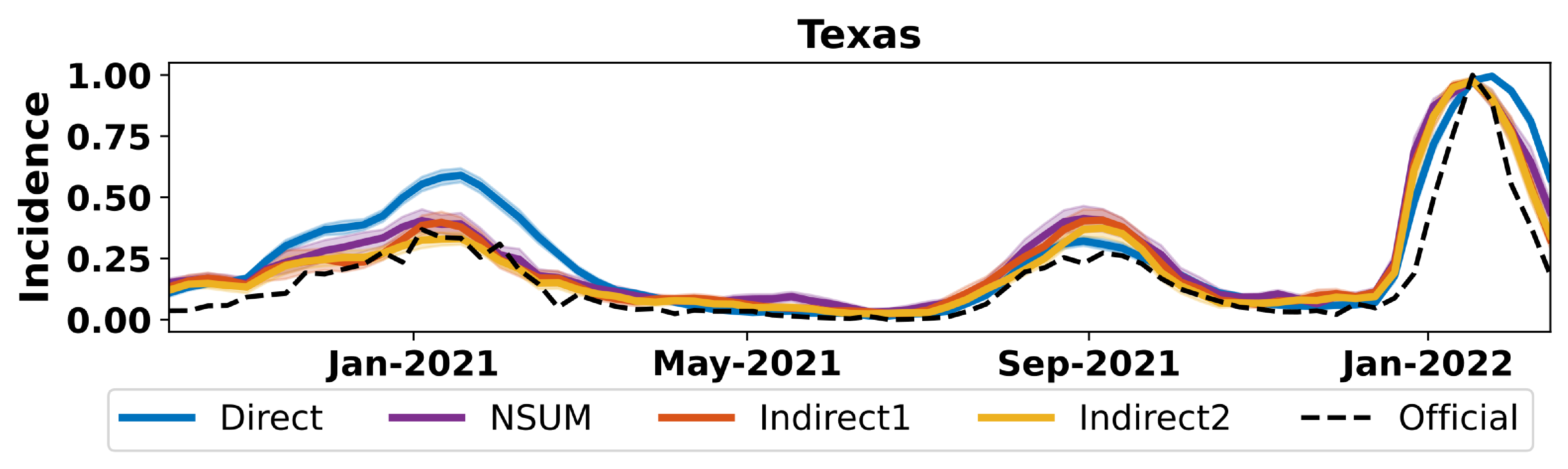} 
    \\
    \hspace{-10pt}
    \includegraphics[width=\linewidth]{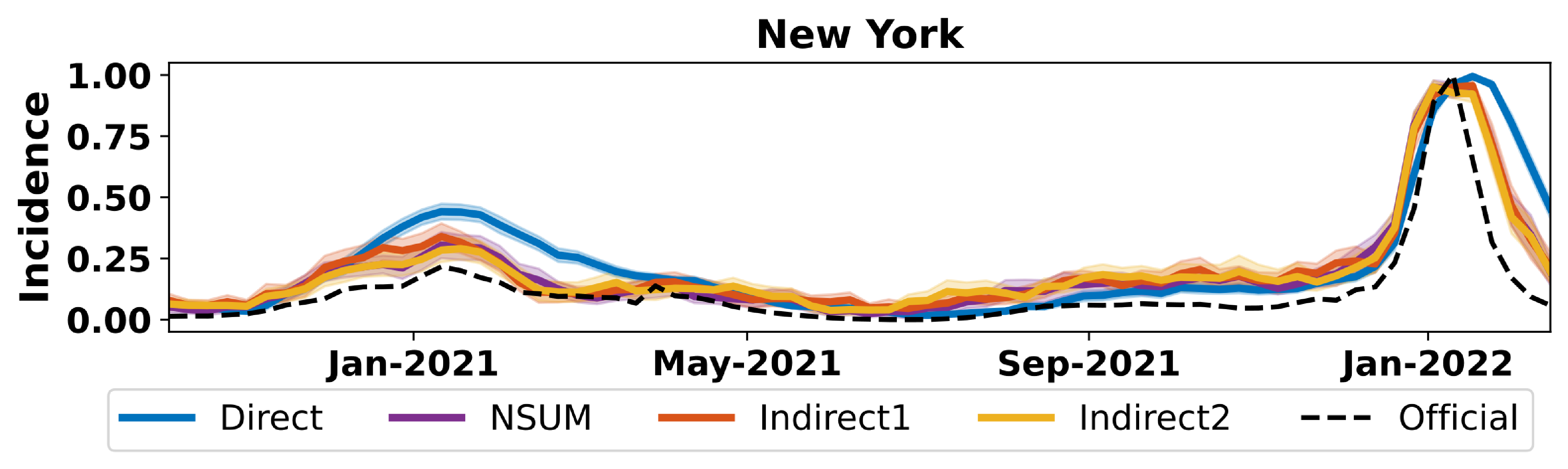} 
    \\
    \hspace{-10pt}
    \includegraphics[width=\linewidth]{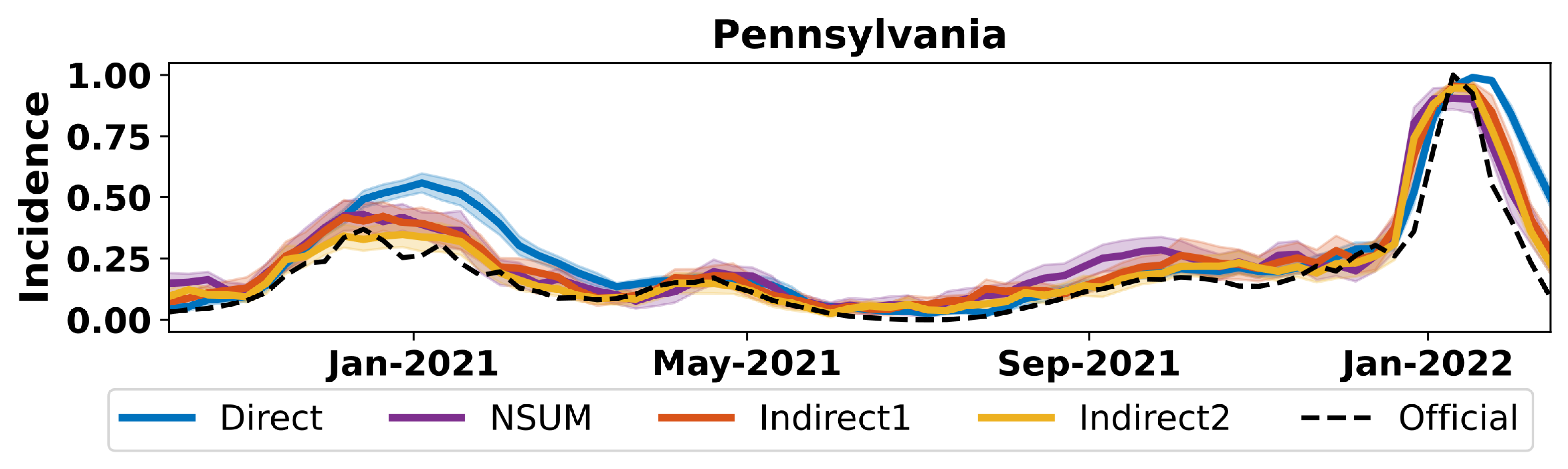}
\end{tabular}
\caption{Normalized COVID-19 incidence curves and the 95\% confidence interval using $3\%$ of the daily samples for the four states and for $accum = 7$ and $w=2$.}\label{fig: CTISsr3}
\end{figure}

\begin{figure}[!t]
\begin{tabular}{c}
\hspace{-10pt}
\begin{minipage}{\linewidth}
\begin{center}
\begin{tabular}{c c c c}
\hspace{-10pt}
\includegraphics[height=0.525\linewidth]{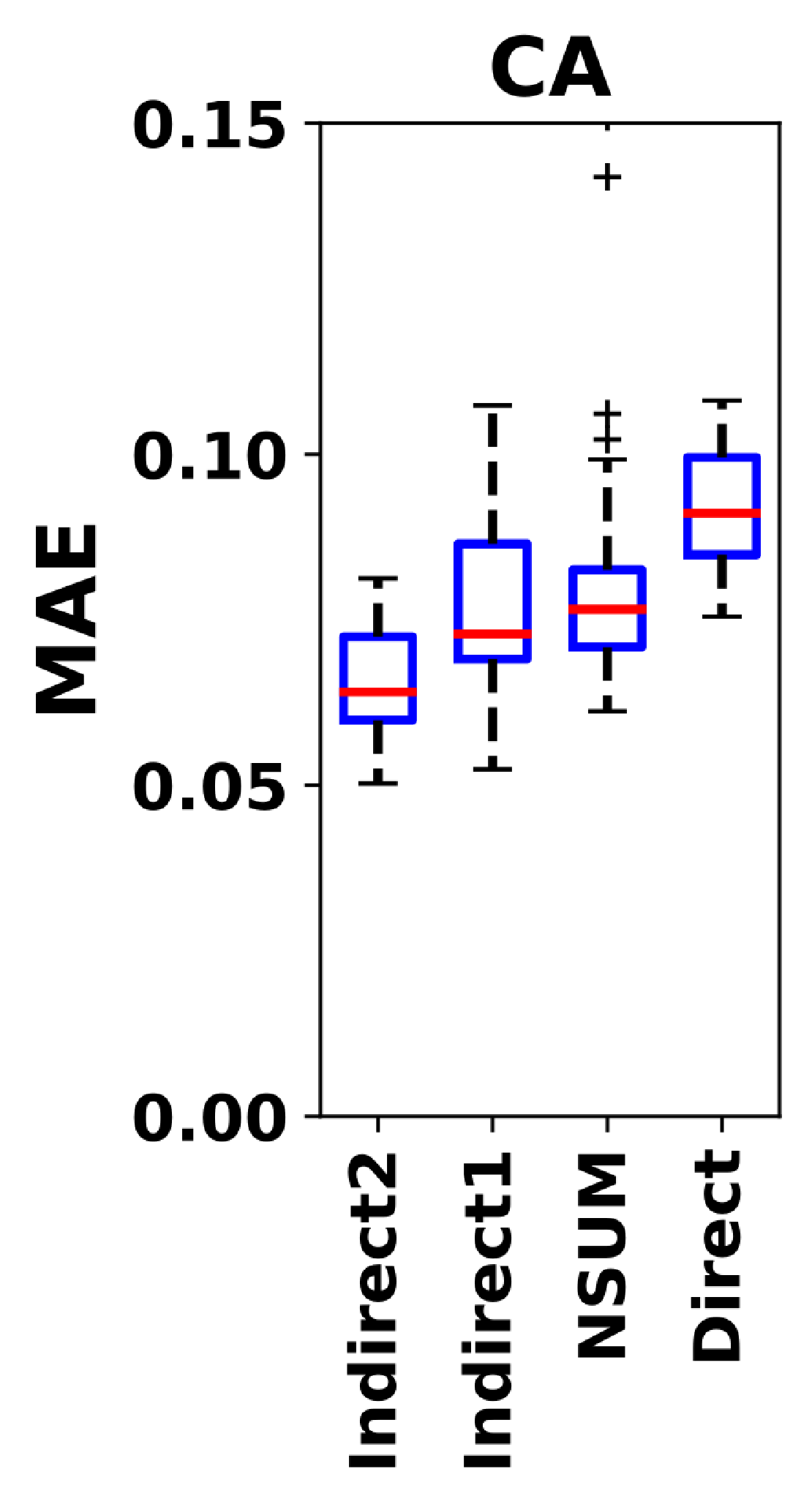}
     &
\hspace{-10pt}
\includegraphics[height=0.525\linewidth]{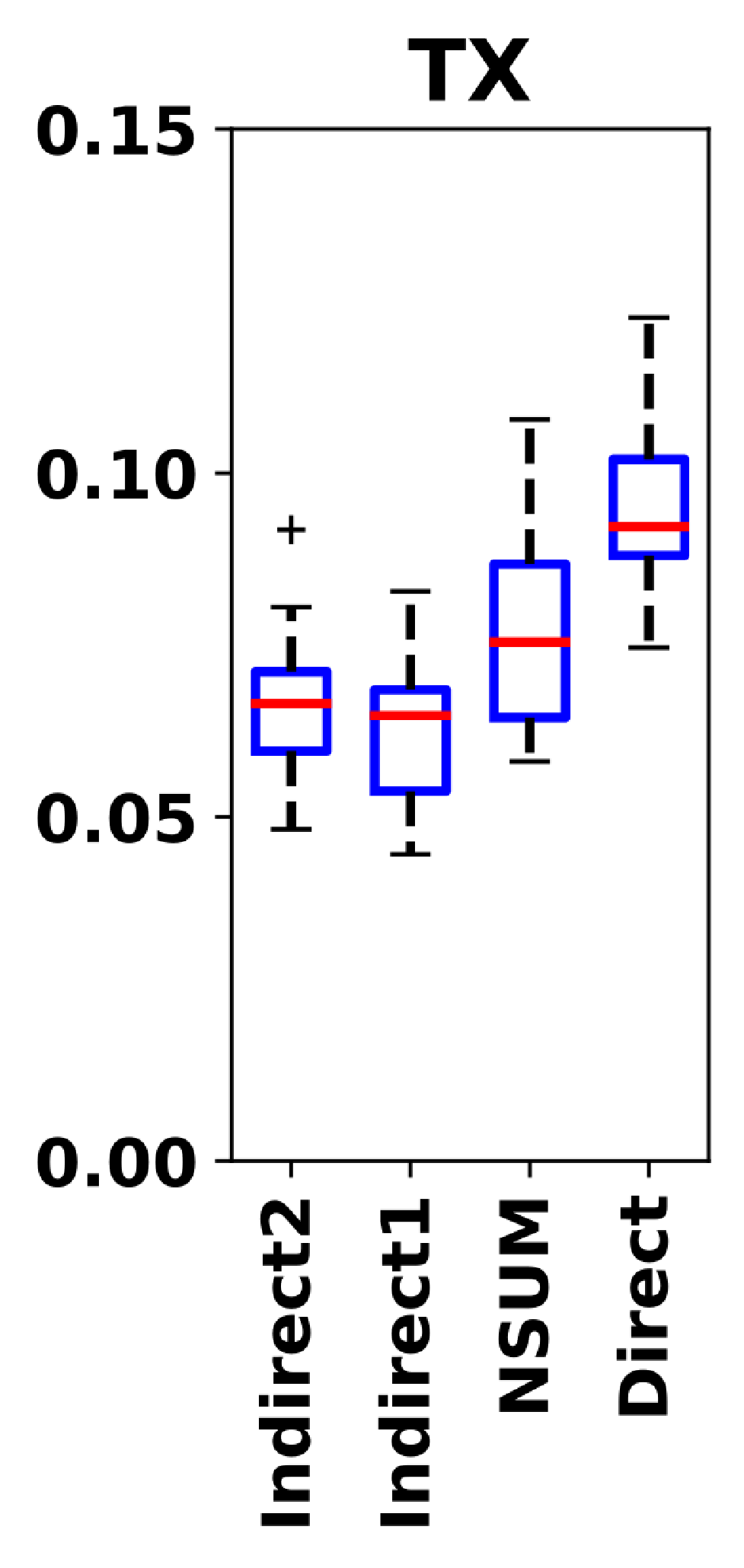}
     & 
\hspace{-10pt}
\includegraphics[height=0.525\linewidth]{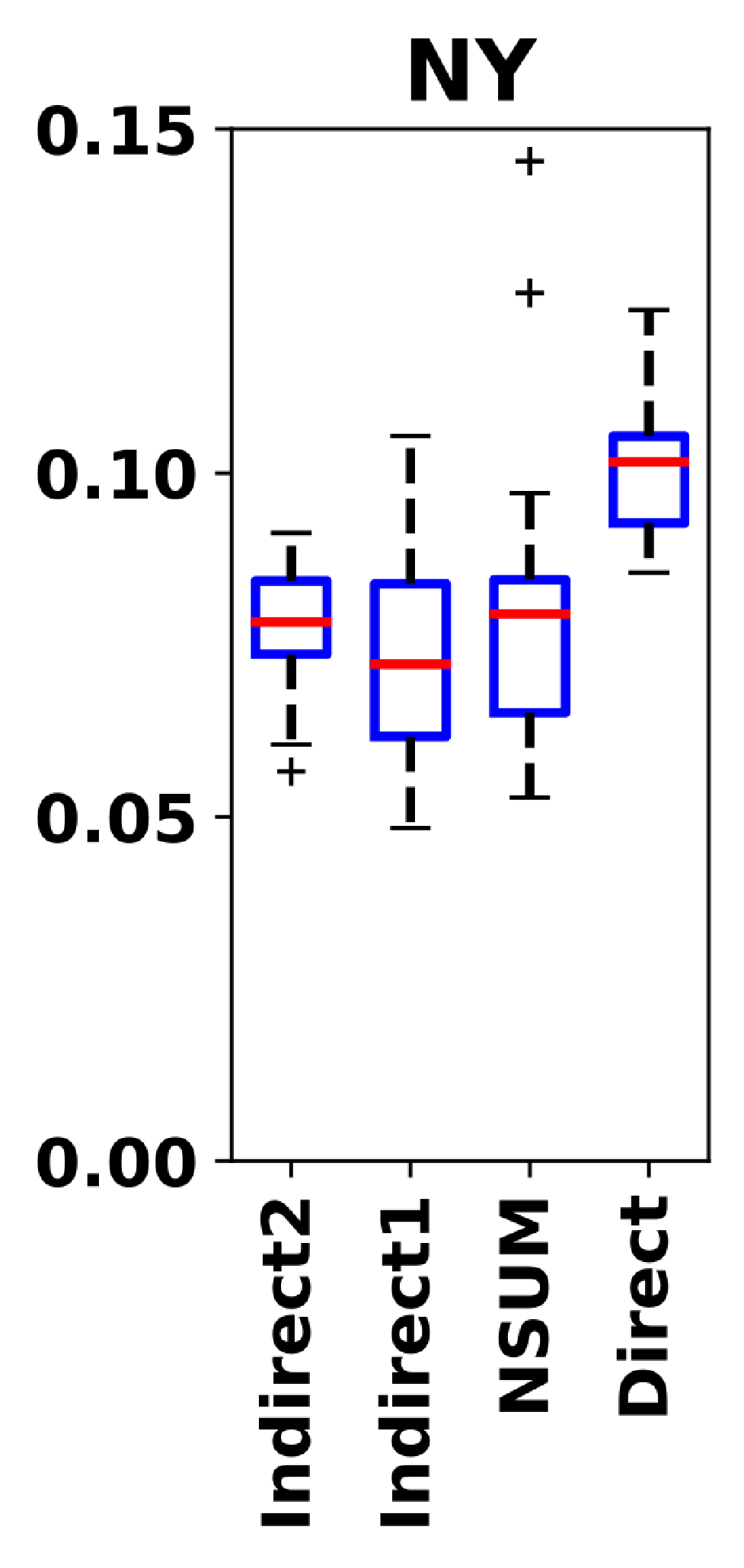}
    &
\hspace{-10pt}
\includegraphics[height=0.525\linewidth]{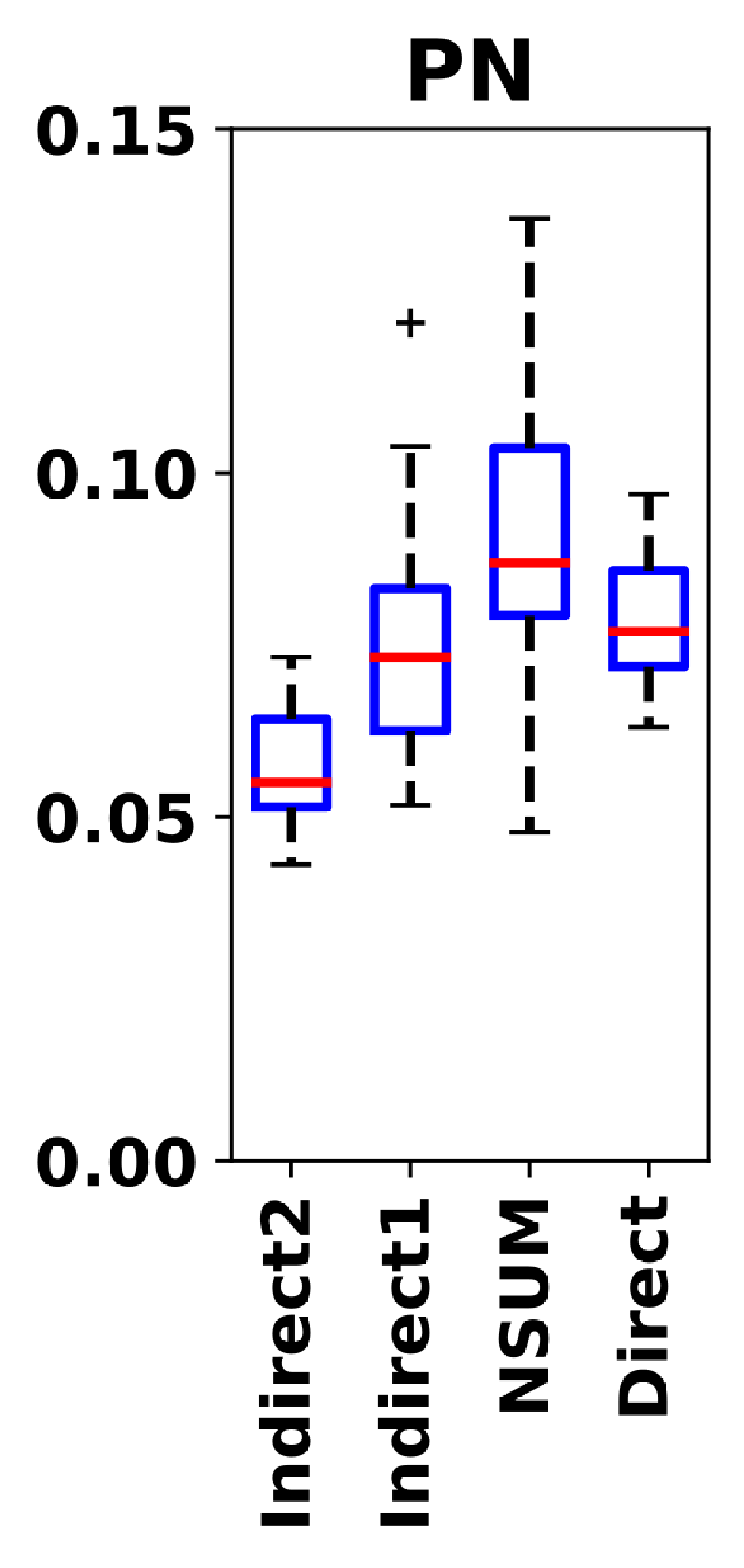}
\end{tabular}
\end{center}
\end{minipage}
\end{tabular}
    \caption{MAE of the normalized COVID-19 incidence curves using $10\%$ of the daily samples for the four states and for $accum=7$ and $w=2$.}
    \label{fig:my_label-7-2_sr10}
\end{figure}

\begin{figure}[!t]
\begin{tabular}{c}
\hspace{-10pt}
\begin{minipage}{\linewidth}
\begin{center}
\begin{tabular}{c c c c}
\hspace{-10pt}
\includegraphics[height=0.525\linewidth]{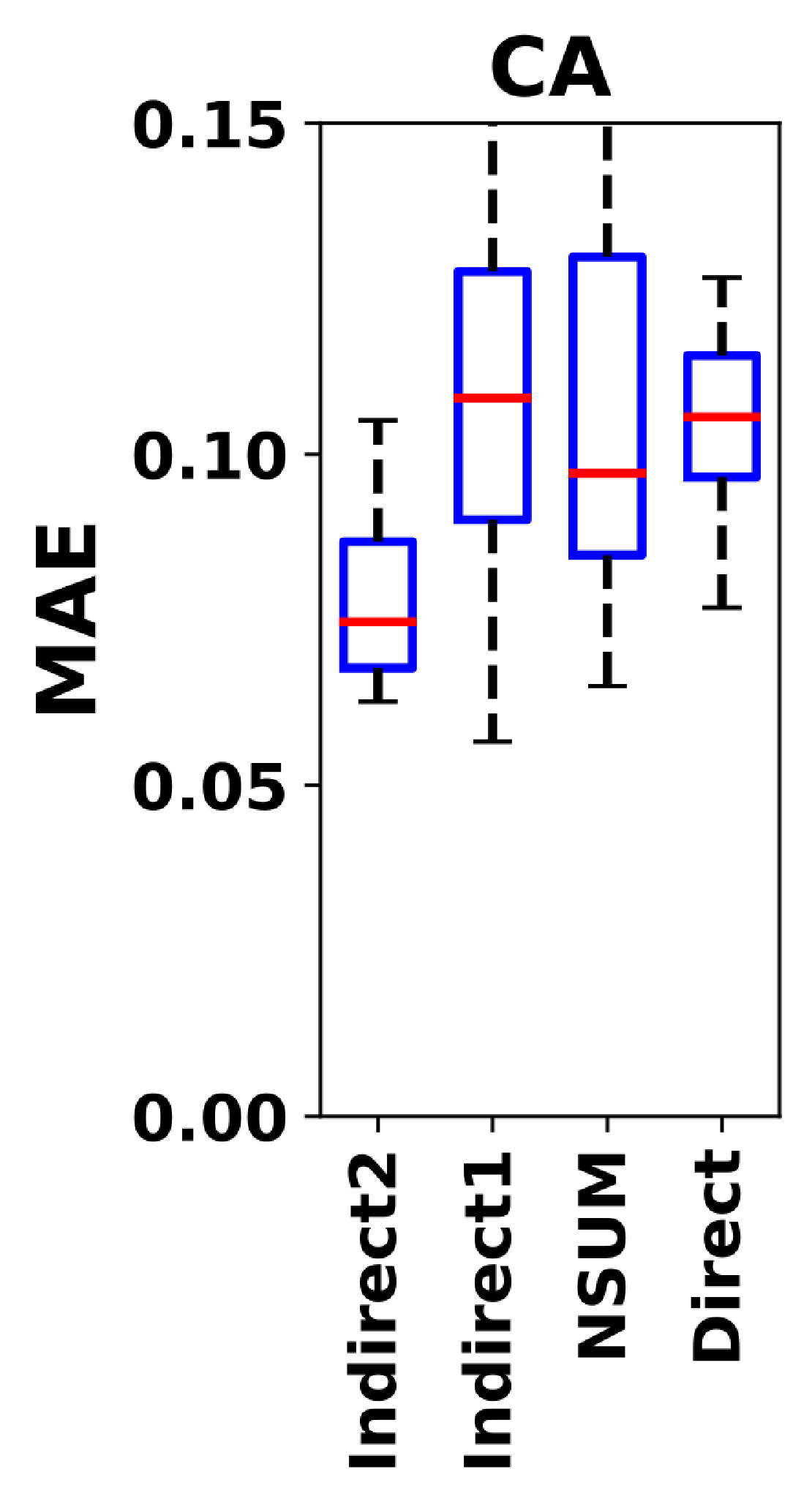}
     &
\hspace{-10pt}
\includegraphics[height=0.525\linewidth]{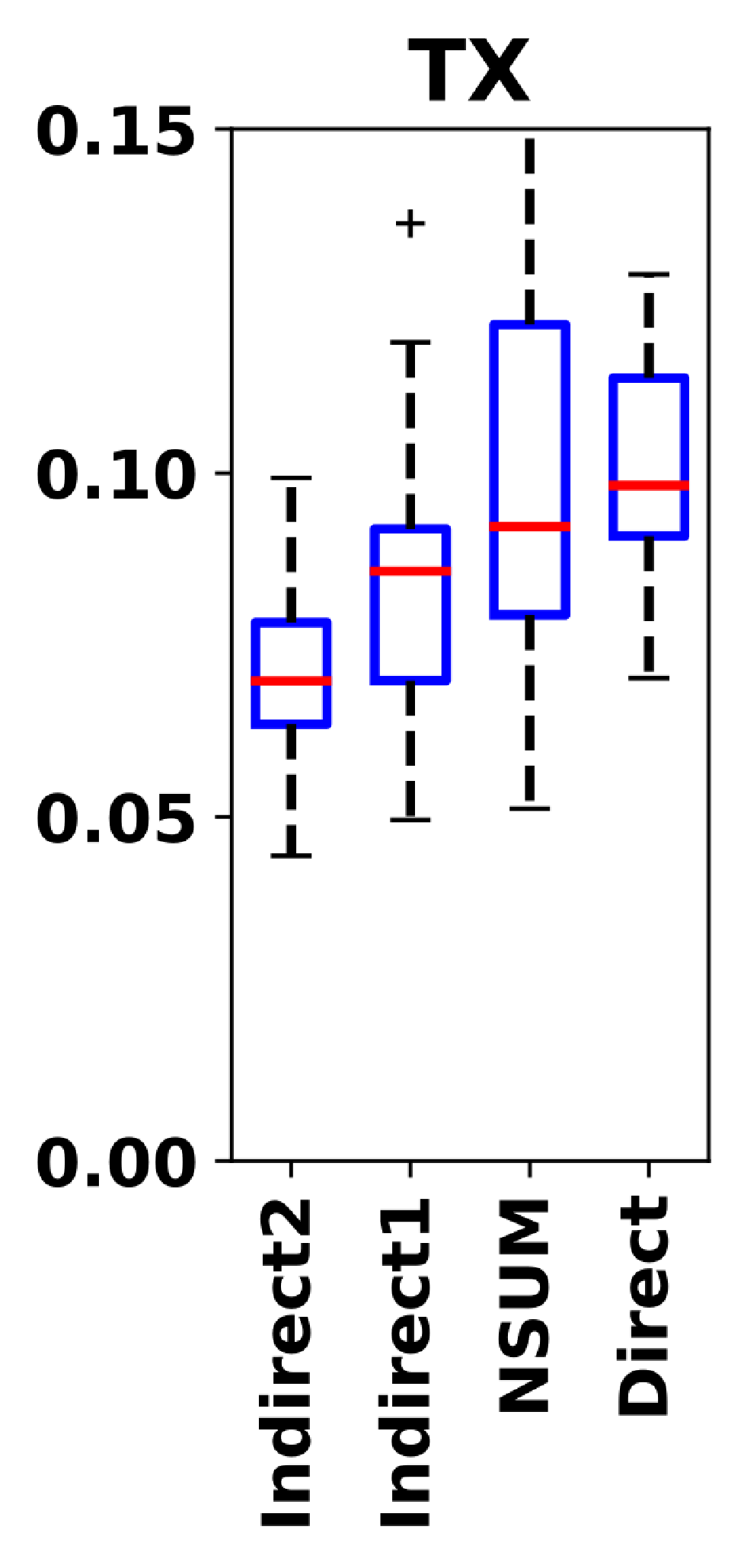}
     & 
\hspace{-10pt}
\includegraphics[height=0.525\linewidth]{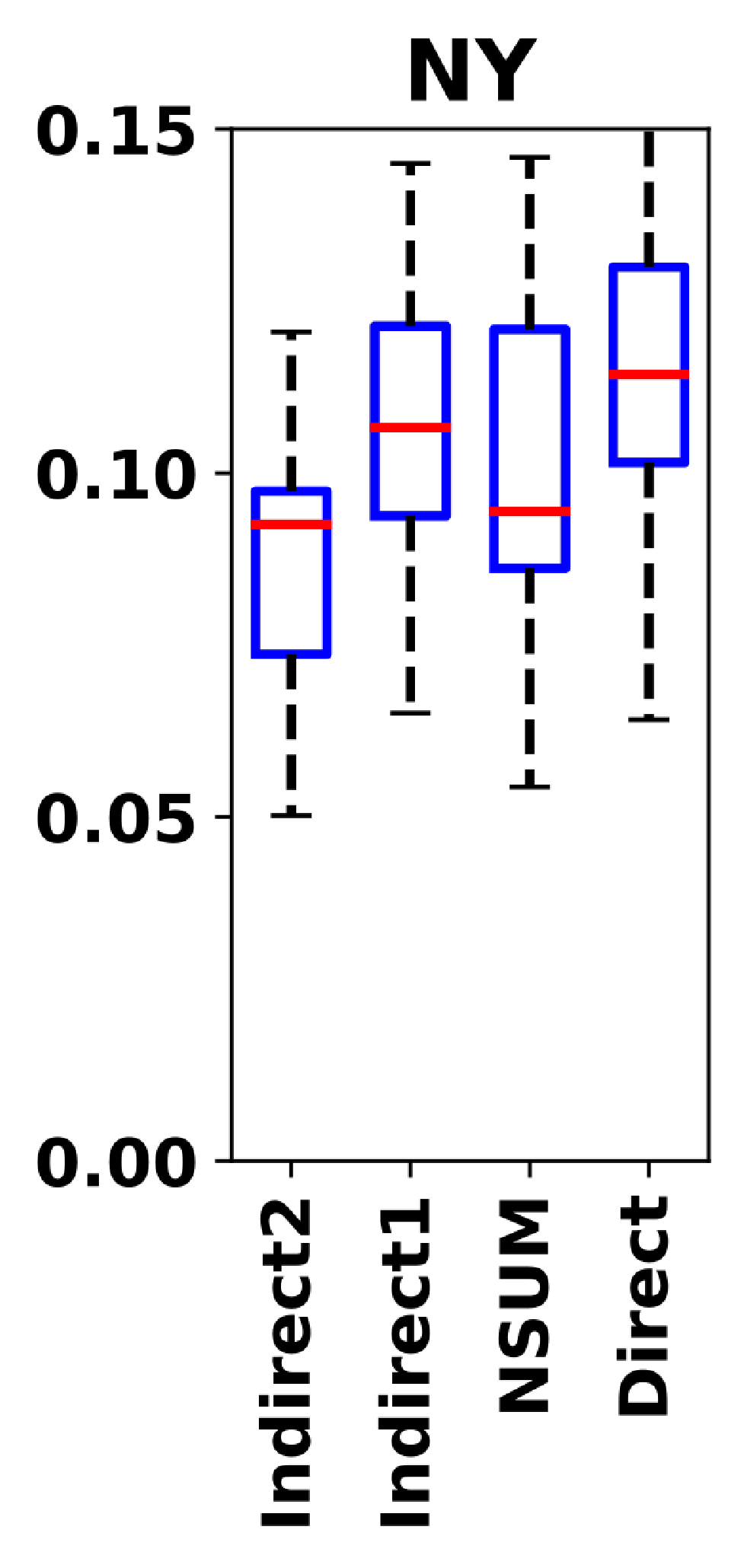}
    &
\hspace{-10pt}
\includegraphics[height=0.525\linewidth]{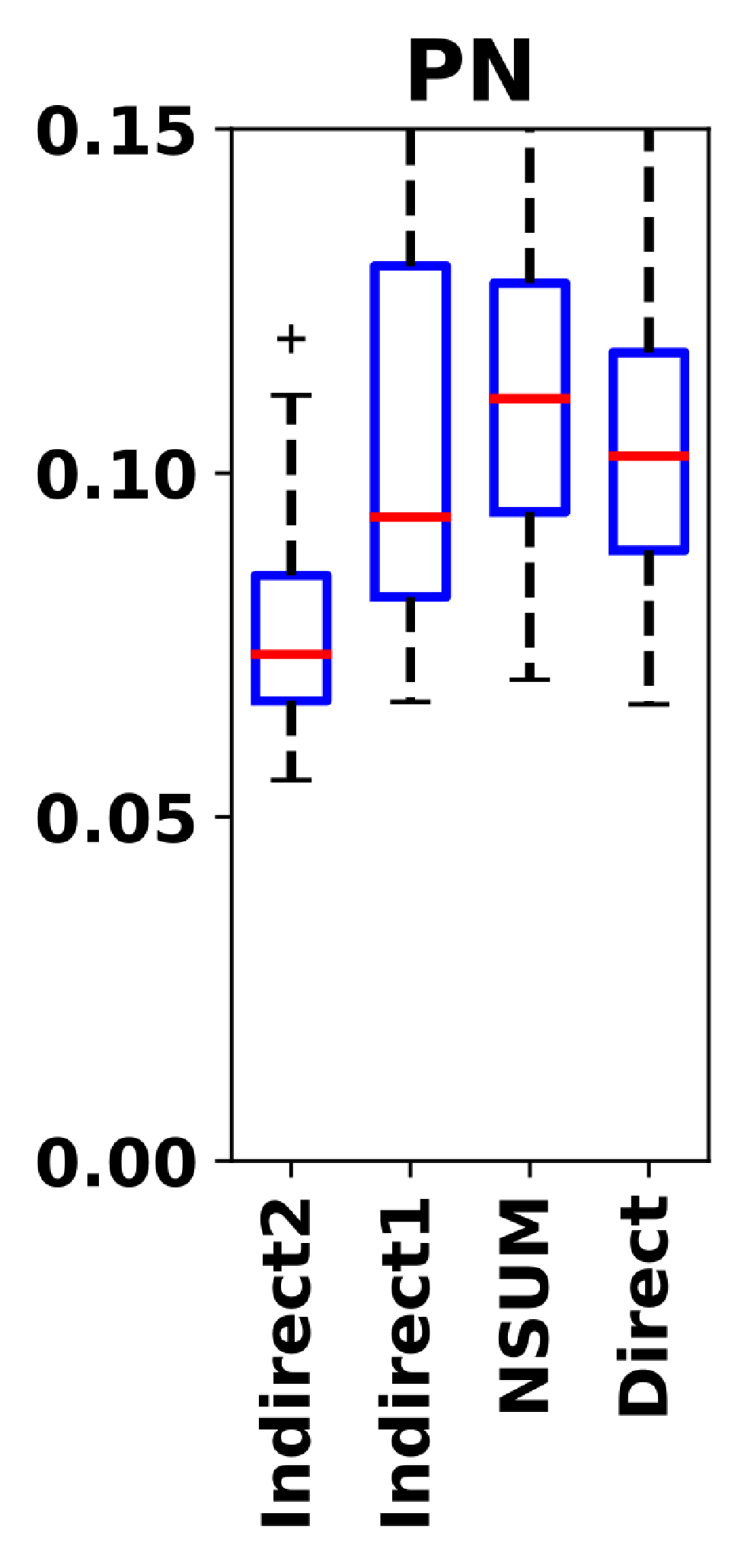}
\end{tabular}
\end{center}
\end{minipage}
\end{tabular}
    \caption{MAE of the normalized COVID-19 incidence curves using $3\%$ of the daily samples for the four states and for $accum=7$ and $w=2$.}
    \label{fig:my_label-7-3_sr3}
\end{figure}

\paragraph{Undersampling Real Data} To demonstrate that our approach works with a small number of responses as well, we performed experiments where we undersampled the US CMU-CTIS data.
Figure \ref{fig: CTISsr10} displays the averaged COVID-19 incidence curves and the corresponding 95\% confidence interval from direct and indirect surveys using 10\% of the daily samples for the four states and for $accum = 7$ and $w=2$. Specifically, the curves are the ensemble average of 25 realizations of the experiment, where a random subset of 10\% of the daily samples was selected at each realization. As can be seen in this figure, the curves from direct and indirect surveys closely follow the trend of the curve obtained from official data. Fig. \ref{fig:my_label-7-2_sr10} illustrates the boxplots of errors obtained by the various methods using 10\% of the daily samples for the four states and for $accum = 7$ and $w=2$. As can be seen in these boxplots, the indirect methods yield the lower median errors compared to those obtained by NSUM and direct techniques. Figures \ref{fig: CTISsr3} and \ref{fig:my_label-7-3_sr3} show, respectively the normalized incidence curves and the distributions errors using 3\% of the daily samples for the four states and for $accum = 7$ and $w=2$.

\paragraph{More Simulation Results}

Here we present the boxplots omitted from the main paper showing the impact of $d$ and $n_d$. These parameters had negligible impact.

\begin{figure}[!t]
    \centering
    \includegraphics[width=\columnwidth,trim={35 0 50 5},clip]{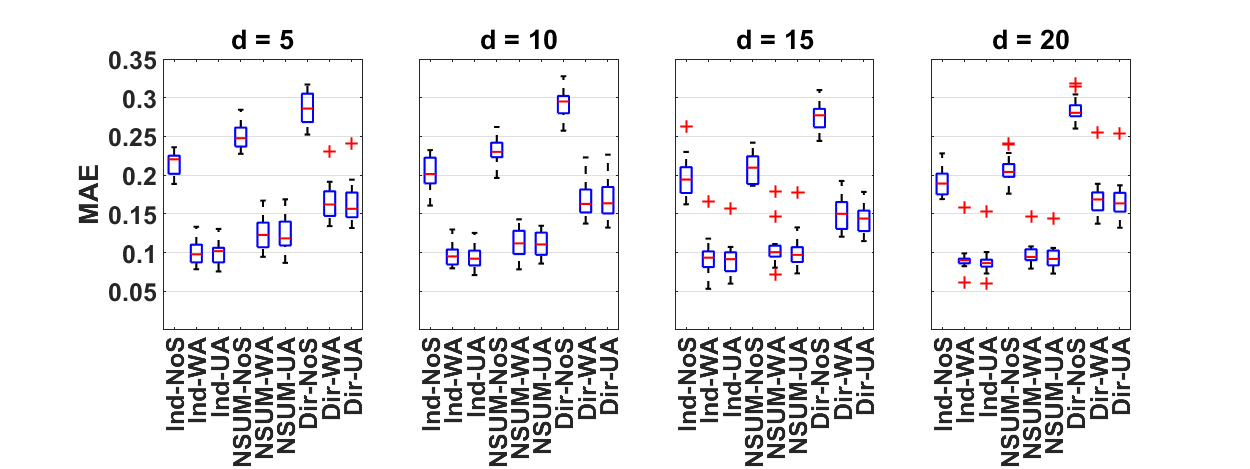}
    \caption{MAE vs $d$, approximate average degree of the respondents. Choice of $d$ does not significantly affect relative performance of the methods}
    \label{fig:ae_d}
\end{figure}
\begin{figure}[!t]
    \centering
    \includegraphics[width=\columnwidth,trim={35 0 50 5},clip]{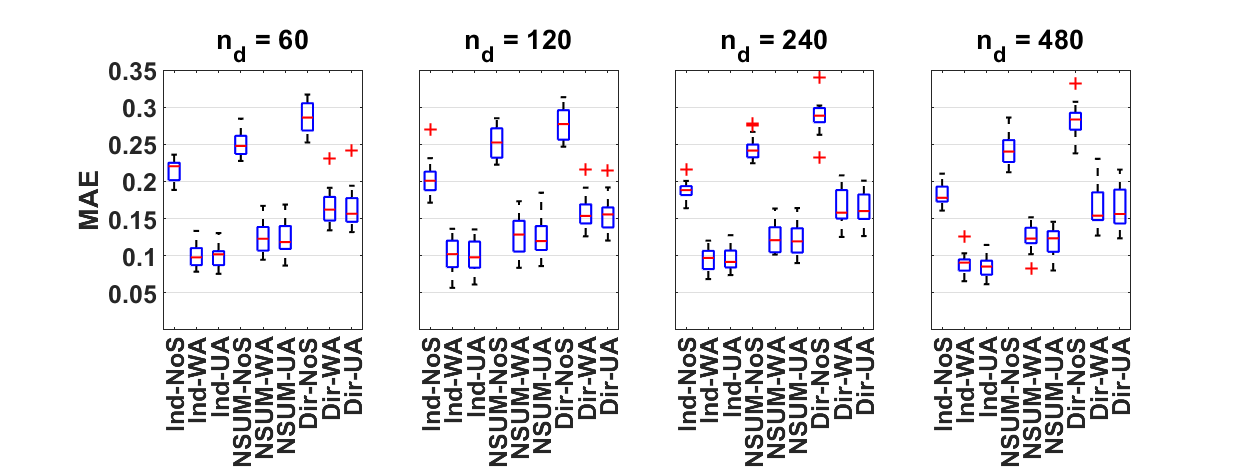}
    \caption{MAE vs $n_d$, the subpopulation potentially covered by the sampled respondents. Choice of $n_d$ does not significantly affect relative performance of the methods}
    \label{fig:ae_n_d}
\end{figure}

\subsection{Results on UMD surveys}
\begin{figure}[!t]
    \centering
    \includegraphics[width=\columnwidth]{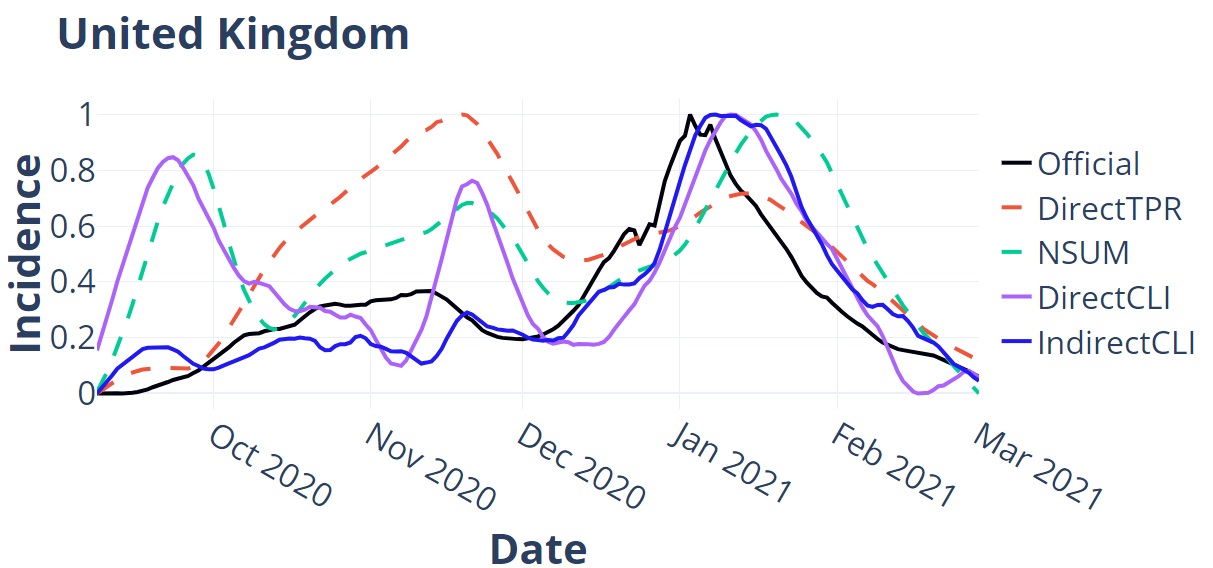}
    \caption{Comparison of estimated incidences from UMD-CTIS surveys. Our approach (IndirectCLI) provides a better estimate than DirectCLI, and even better than asking directly about positive tests (as opposed to indirect on CLI).}
    \label{fig:UK}
\end{figure}

We also conducted experiments on UMD surveys~\footnote{\url{https://covidmap.umd.edu/}} that collected survey data at the country level for various countries. In this dataset, there are responses regarding if one has tested positive recently (DirectTPR), if they have COVID-like symptoms (DirectCLI), and the ``indirect'' question -- how many they know who have COVID-like symptoms. The last question allows us to perform estimation using our approach (Indirect). We used $w=1$, i.e., a window of $3$ to aggregate the responses. There is no information on the personal network size which is needed for NSUM. Therefore, we used an existing technique\footnote{\url{https://arxiv.org/abs/2108.03284}} that estimates the personal network size for comparison. The ground truth was obtained from the reported case count available from Our World in Data\footnote{\url{https://ourworldindata.org/covid-cases}}.
The results are presented in Table~\ref{tab:umd}. We note that Indirect outperforms DirectCLI and NSUM on all countries with a significant margin in most. In fact, Indirect (regarding CLI) provides a better estimate than directly asking about positive tests in most countries. Figure~\ref{fig:UK} shows the comparison of estimated incidences for the United Kingdom. Our approach (IndirectCLI) most closely resembles the ground truth. Our estimates are not only better than DirectCLI but also better than asking directly about positive tests (as opposed to indirect on CLI).

\begin{table}[!t]
    \centering
    \begin{tabular}{|l|c|ccc|}
    \hline
        Country & DirectTPR & NSUM & DirectCLI & Indirect \\
        \hline
Brazil  & \textbf{0.14}      & 0.17 & 0.15      & 0.15     \\
Canada  & 0.14      & 0.19 & 0.4       & \textbf{0.12}     \\
France  & 0.17      & 0.24 & 0.22      & \textbf{0.13}     \\
Germany & \textbf{0.1}       & 0.19 & 0.29      & 0.12     \\
Israel  & 0.32      & 0.19 & 0.17      & \textbf{0.12}     \\
Italy   & 0.17      & 0.13 & 0.05      & \textbf{0.1}      \\
Japan   & 0.15      & 0.23 & 0.25      & \textbf{0.08}     \\
Spain   & 0.31      & 0.2  & 0.16      & \textbf{0.1 }     \\
Turkey  & 0.24      & 0.29 & 0.16      & \textbf{0.12}     \\
UK      & 0.24      & 0.26 & 0.23      & \textbf{0.11}     \\
\hline
    \end{tabular}
    \caption{MAE on country-level comparison from UMD-CTIS surveys.}
    \label{tab:umd}
\end{table}

\section{Additional Methodology Details}

\subsection{Code}

All our code and aggregate datasets are publicly available\footnote{Our code is available at \url{https://github.com/GCGImdea/coronasurveys/tree/master/papers/2024-AAAI-Nowcasting-Temporal-Trends-Using-Indirect-Surveys}.}.
The synthetic experiments were run on a machine with Intel(R) Core(TM) i8 CPU, 3GHz, 6 cores, and 16GB RAM. The code was written in MATLAB 2021b. The  experiments on real-world data were conducted on a machine with Intel(R) Xeon(R) CPU, 2.60GHz, 16 cores, and 32GB RAM. The pre-processing code to generate aggregate data from CTIS survey was written in R. The results demonstarted in the paper were generated using Python.

\subsection{Outlier Filtering}

In order to remove inconsistent and potentially malicious responses to the CTIS survey we apply a simple outlier filter based on the following rules:
\begin{itemize}
    \item We remove responses that report negative values in numeric questions that can only have non-negative responses.
    \item We remove responses that report more than 100 in the number of people in household with COVID-like illness (CLI), in the number of people in the household in any age range ($<18$, 18-64, and $\geq 65$), or in the number of days with symptoms.
    \item We remove inconsistent responses that report people in household with CLI but no symptoms, or report more people with CLI than the household size.
\end{itemize}

\subsection{Synthetic Experiments}

Here we present more details on the simulations.

\paragraph{Epidemic simulation}
We use a simple SIR model to simulate an epidemic with varying infection parameters whose reproduction number R0~\cite{dietz1993estimation} is above 2. At random points in time, we introduce ``interventions'' that reduce R0 smoothly to a value below 1. We run several such simulations and pick one that produces multiple peaks over 600 days, to emulate realistic epidemics like Influenza and COVID-19 that have multiple waves. We acknowledge that, in reality, not all infections will be detectable. However, assuming that each infection will be detected with a fixed probability only scales the time-series $I(t)$ (the number of infectious individuals by the SIR process) by a constant. Therefore, for the purpose of this study, we directly use $I(t)$ to compute the hidden population over time.

\paragraph{Survey simulation.}

We simulate, at each time $t$, random sets $L$ and $R$ of sizes $n$ and $n_d$, and a bipartite digraph between them with terminal set $L$, which represents the surveyed individuals. To construct the digraph, we pick the degree of vertex $u\in L$, $\delta_u$ from a power law distribution $p_k \propto k^{-2}, k\in [1, n_d/2]$, such that mean degree is approximately $d$. The choice of the probability distribution was driven by the intention to introduce some skewness in the degree distribution. Then, we add an edge between $u$ and $v \in R$ with a probability $\delta_u/n_d$.

For the indirect surveys simulation, we randomly infect each node in $R$ with the probability $\sum_{\tau=0}^{\mbox{\textit{period}}-1} I(t-\tau)$, where period is the time window within which the respondents are to count the hidden population. For comparison, we also introduced the traditional NSUM approach \cite{killworth1998social}, where the response from each node is normalized by its degree.
For direct surveys, we infect nodes in $L$ itself and note the number of infections produced in $L$.

\section{Algorithm in Practice}
\label{sec:alg}
\algtext*{EndWhile}
\algtext*{EndIf}
\begin{algorithm}
\caption{Indirect Survey Estimation}\label{alg:cap}
\begin{algorithmic}
\Procedure{AggregatedEstimate}{$\{\bar{X}_t\}, \{n_t\}, \lambda$}
\State Pick some $w$
\While{$w>0$}
\State Compute $\sigma_n$ and $\mu_n$ for the window $w$
\State $\lambda_1 \gets$  RHS of Inequation~\ref{eqn:thm_accum1}
\State $\lambda_2 \gets$ RHS of Inequation~\ref{eqn:thm_accum2}
\If{$\min\{\lambda_1, \lambda_2\} \leq \lambda$}
    \State Break
\EndIf
\State $w \gets w - 1$
\EndWhile
\Return $\bar{X}_{t, w} \gets \frac{\sum_{i=-w}^w n_{t+i} \bar{X}_{t+i}}{\sum_{i=-w}^w n_{t+i}}$
\EndProcedure
\end{algorithmic}
\end{algorithm}
Suppose we have some bounds on $\epsilon_{f, 1}$ and $\epsilon_{f, 2}$. This could be obtained from prior experience. For example, if we are estimating the number of people who test positive for COVID-19, we can use prior trend of COVID-19 when the reporting quality was better (before home tests). 
We can use these along with Theorems~\ref{thm:accum} and~\ref{thm:accum2} to identify a window that is likely to improve our estimation as given in Algorithm~\ref{alg:cap}. Suppose we are given a threshold $\lambda$ acceptable fractional deviation, e.g., $\lambda = 0.1$. Then we can search for a $w$ that either satisfies Inequality~\ref{eqn:thm_accum1} or Inequality~\ref{eqn:thm_accum2}. If Inequality~\ref{eqn:thm_accum1} is satisfied, then by Theorem~\ref{thm:accum}, aggregating with window $w$ must be better than no smoothing. If Inequality~\ref{eqn:thm_accum2}, then the same applies due to Theorem~\ref{thm:accum2}.
We can apply this aggregation over discrete bins (daily, weekly, monthly) or in a rolling fashion (smoothing), or some combination of the two.

\end{document}